\documentclass[12pt,a4paper]{article}

\usepackage{latexsym}
\usepackage{amsmath}
\usepackage{amsfonts}
\usepackage{amssymb}

\newtheorem{theorem}{Theorem}[section]

\newtheorem{proposition}[theorem]{Proposition}

\newtheorem{lemma}[theorem]{Lemma}

\newcommand{\qedsymb}{{\em Q.E.D.}}

\newcommand{\be}{\begin{equation}}
\newcommand{\ee}{\end{equation}}
\newcommand{\bea}{\begin{eqnarray}}
\newcommand{\eea}{\end{eqnarray}}

\renewcommand{\theequation}{\arabic{section}.\arabic{equation}}

\def\ad{{\mathrm{ad}}}                  %
\def\ri{{\mathrm{i}}}                   %
\def\rr{{\mathrm{r}}}                   %
\def\bR{{\mathbb R}}                    %
\def\bC{{\mathbb C}}                    %
\def\bT{{\mathbb T}}                    %
\def\1{{\mbox{\boldmath $1$}}}          %
\def\0{{\mbox{\boldmath $0$}}}          %
\def\red{\mathrm{red}}                  %
\def\cF{{\cal F}}                       %
\def\cR{{\cal R}}                       %
\def\cM{{\mathcal M}}                   %
\def\cH{{\mathcal H}}                   %
\def\bR{{\mathbb R}}                    %
\def\bZ{{\mathbb Z}}                    %
\def\ri{{\mathrm{i}}}                   %
\def\tr{\mathrm{tr}}                    %
\def\diag{\mathrm{diag}}                %
\def\mQ{{Q}}                            %
\newcommand{\dd}{\mathrm{d}}            %
\newcommand{\Inv}{\mathrm{Inv}}         %
\newcommand{\Lie}{\mathrm{Lie}}         %
\newcommand{\Id}{\mathrm{Id}}           %
\newcommand{\Span}{\mathrm{span}}       %
\newcommand{\End}{\mathrm{End}}         %
\newcommand{\cB}{\mathcal{B}}           %
\newcommand{\cA}{\mathcal{A}}           %
\newcommand{\cY}{\mathcal{Y}}           %
\newcommand{\cG}{\mathcal{G}}           %
\newcommand{\cK}{\mathcal{K}}           %
\newcommand{\cJ}{\mathcal{J}}           %
\newcommand{\cW}{\mathcal{W}}           %
\newcommand{\bfPi}{\mathbf{\Pi}}        %
\newcommand{\bfI}{\mathbf{I}}           %
\newcommand{\bfq}{\mathbf{q}}           %
\newcommand{\mfsu}{\mathfrak{su}}       %
\newcommand{\mfu}{\mathfrak{u}}         %
\newcommand{\mfgl}{\mathfrak{gl}}       %
\newcommand{\mfsl}{\mathfrak{sl}}       %
\newcommand{\ket}[1]{| #1 \rangle}      %
\begin{document}

\vspace*{0.5cm}
\begin{center}
{\Large \bf Derivations of the trigonometric $BC_n$ Sutherland model
by quantum Hamiltonian reduction}
\end{center}

\vspace{0.2cm}

\begin{center}

 L.~Feh\'er${}^{a,b}$ and B.G.~Pusztai${}^{c}$  \\

\medskip
\bigskip

${}^{a}$Department of Theoretical Physics, MTA  KFKI RMKI\\
H-1525 Budapest, P.O.B. 49,  Hungary \\
e-mail: lfeher@rmki.kfki.hu

\bigskip

${}^b$Department of Theoretical Physics, University of Szeged\\
Tisza Lajos krt 84-86, H-6720 Szeged, Hungary\\

\bigskip

${}^c$Bolyai Institute, University of Szeged\\
Aradi v\'ertan\'uk tere 1, H-6720 Szeged, Hungary\\
e-mail: gpusztai@math.u-szeged.hu

\end{center}

\vspace{0.2cm}

\begin{abstract}

The $BC_n$ Sutherland Hamiltonian with
coupling constants parametrized by three arbitrary integers
is derived by reductions of the Laplace operator of the group $U(N)$.
The reductions are obtained by applying the Laplace operator on spaces of certain vector
valued functions equivariant under suitable symmetric subgroups of $U(N) \times U(N)$.
Three different reduction schemes are considered, the simplest one being
the compact real form of the reduction  of the Laplacian of
$GL(2n,\bC)$ to the complex
$BC_n$ Sutherland Hamiltonian previously studied by Oblomkov.

\end{abstract}

\newpage

\section{Introduction}
\setcounter{equation}{0}

The family of Calogero-Sutherland type  many-body models is
very important both in physics and mathematics, as is amply demonstrated in the
reviews \cite{OPRep,Heck,SutWSci,Sas,Eting,Poly}.
In this paper we focus on the group theoretic derivation of the
trigonometric Sutherland models introduced
by Olshanetsky and Perelomov \cite{OPInv} in correspondence with the
 crystallographic root systems.
The Hamiltonian of the model  associated with the roots system  $\cR$ is given by
\be
H_{\cR} =- \frac{1}{2} \Delta + \frac{1}{4}\sum_{\alpha\in \cR}
\frac{\vert \alpha \vert^2 \mu_{\alpha}(\mu_\alpha + 2 \mu_{2\alpha} -1)}{\sin^2(\alpha \cdot q)},
\label{1.1}\ee
where $\Delta$ is the Laplacian on the Euclidean space of the roots and the
$\mu_{\alpha}$ are arbitrary real constants depending only on the lengths of the roots, with
$\mu_{2\alpha}:=0$ if $2\alpha \notin \cR$.
In the original $A_{n-1}$ case the model was solved by Sutherland \cite{Sut72}.
An interesting general observation \cite{OPfa} is  that the radial
part of the
Laplace operator of any compact Riemannian symmetric space is always conjugate to a
Sutherland operator (\ref{1.1}) built on
the root system of the symmetric space,
with coupling constants determined by the multiplicities of the roots.
This observation showed  the algebraic integrability of the  resulting Hamiltonians $H_\cR$ at
(small) finite sets of coupling
constants  and  inspired later developments.
The integrability, and  exact solvability in terms of a triangular structure,
was first established for the models (\ref{1.1}) in full
generality  by Heckman and Opdam \cite{HeOp,Opd}.
Their technique is based on differential-reflection operators  belonging
to the Hecke algebraic generalization  of harmonic analysis \cite{Heck,Cher}.

The Hecke algebraic approach is very powerful, but it is still desirable
to treat as many cases of the models (\ref{1.1}) in  group theoretic terms as possible.
Important progress in this direction was achieved  by Etingof, Frenkel and Kirillov \cite{EFK}
who worked out the quantum mechanical version of the classical Hamiltonian
reduction due to Kazhdan, Kostant and Sternberg \cite{KKS} and thereby
showed that the $A_{n-1}$  Sutherland Hamiltonian arises as the restriction of
the Laplace operator of $SU(n)$ to certain vector valued spherical functions.
A spherical function $F$ on $SU(n)$ with values in the $SU(n)$ module $V$ satisfies
the equivariance condition $F(g x g^{-1}) = g\cdot F(x)$ and thus it is uniquely
determined by its restriction to the maximal torus $\bT<SU(n)$. It is easily seen that the
restricted function $f=F\vert_{\bT}$ must vary in the zero-weight subspace
$V^{\bT}$ and the action of  the Laplace operator of $SU(n)$ on $F$
can be expressed by the action of a scalar differential operator on $f$
whenever $\dim(V^\bT)=1$.
This latter condition singles out the symmetric tensorial powers
$V = S^{kn}(\bC^n)$ ($k\in \bZ_{\geq 0}$) and their duals  among the irreducible highest weight
representations of $SU(n)$, and the resulting scalar differential operator turns out
to be the Sutherland operator $H_{A_{n-1}}$ with coupling parameter $\mu_\alpha=k+1$.

The above arguments cannot be extended to the simple Lie groups beyond $SU(n)$,
since in general
they do not admit
non-trivial highest weight representations with multiplicity one for the zero
weight\footnote{
The only
exceptions \cite{Ber,Howe} are the defining representation of $SO(2n+1)$
and the $7$-dimensional representation of $G_2$.
In the former case we have checked that
the reduced Laplacian gives a decoupled system.}.
However, taking any
compact connected Lie group $Y$,  there exist other nice actions of certain subgroups
of $Y \times Y$ on $Y$ for which one can try to generalize the above arguments.
Indeed \cite{HPTT}, if $G$ is the fixed point set of an involution of $Y \times Y$, then
every orbit of the  natural  action of $G$ on $Y$ can be intersected by a toral subgroup
$A<Y$. Therefore the $G$-equivariant functions on $Y$ with values in a representation $V$ of $G$
give rise to $V^K$-valued functions on $A$, where $K$ is the isotropy group of the
generic elements of $A$.
Moreover, if $\dim(V^K)=1$, then the application of the Laplace
operator of $Y$ on $C^\infty(Y,V)^G$ may induce a scalar Sutherland operator.
The group actions just alluded to are called Hermann actions.
They received a lot of attention in differential geometry
(see e.g.~\cite{HPTT,Kolr} and references therein), but their use
for the construction of integrable systems still has not been explored systematically.

The goal of this paper is to explain
that certain Hermann actions on $Y=U(N)$
permit derivations of the $BC_n$ Sutherland Hamiltonian
from the  Laplacian  of   $U(N)$.
The derivations that we present  are partly   motivated by an earlier derivation
found in the complex holomorphic setting in \cite{Obl}, and
by our previous paper  \cite{LMP} where we
discussed how the classical mechanical version of the
trigonometric $BC_n$ model
with three arbitrary coupling constants can be obtained by reducing
the free particle moving on  the group $U(N)$.
Taking for $\cR$ the root system
\be
BC_n= \bigl\{ \epsilon_i \pm \epsilon_j,\, \pm \epsilon_k,\, \pm 2 \epsilon_k\,\vert\,
i,j,k\in\{1,\dots,n\}, \,\, i\neq j\bigr\},
\ee
with orthonormal vectors $\{ \epsilon_i\}$, and
 introducing new coupling parameters $a, b, c$ by the definition
\be
\mu_{\epsilon_i \pm \epsilon_j}:= a+1,
\quad
\mu_{\epsilon_k}: = b-c, \quad
\mu_{2\epsilon_k}: = c + \frac{1}{2},
\label{***}\ee
the Hamiltonian (\ref{1.1}) reads
\be
H_{BC_n} =
-\frac{1}{2} \sum_{j = 1}^n \frac{\partial^2}{\partial q_j^2}
+ \sum_{1 \leq k < l \leq n} \left(
\frac{a(a+1)}{\sin^2(q_k - q_l)} + \frac{a(a+1)}{\sin^2(q_k + q_l)} \right) +
\frac{1}{2}\sum_{j = 1}^n \frac{b^2- \frac{1}{4}}{\sin^2(q_j)}
+ \frac{1}{2}\sum_{j = 1}^n \frac{c^2-\frac{1}{4}}{\cos^2( q_j)}.
\label{**}\ee
In fact, we shall obtain this Hamiltonian
with  \emph{arbitrary non-negative integers $a$, $b$ and $c$}
as  a reduction of the Laplace operator of $U(N)$.
More precisely, we shall present 3 different derivations, for which  $N=2n$, $N=2n+1$ or $N=2n+2$.

There is considerable conceptual overlap between this paper and the above-mentioned work \cite{Obl}
of Oblomkov,
who related the eigenfunctions of the holomorphic $BC_n$ Sutherland operator to vector valued
spherical functions on the group $GL(N,\bC)$.
If we replace $GL(N,\bC)$ by $U(N)$,
then  Oblomkov's  construction leads to our
construction in the most important $N=2n$ case.
However, there are also different cases considered in \cite{Obl} and in this paper even after
such replacement, and the language and the techniques used are rather different.
In fact, we shall obtain the results  by applying a recently developed general framework
of quantum Hamiltonian reduction under polar group actions \cite{TMP}.
We shall raise interesting open questions, too, and
to facilitate  their future investigation we  describe our analysis
in a  self-contained manner.

The organization of the article is as follows.
In the next section  we recall  the necessary notions and results concerning
 quantum Hamiltonian
reductions of
the Laplace operator on a Riemannian manifold that admits
generalized polar coordinates
 adapted to the symmetry group in the sense of \cite{PT}.
In section 3 we specialize to Hermann actions on a compact Lie group $Y$,
and  describe those  Hermann actions on
$Y=U(N)$ that are expected to lead to $BC_n$  Sutherland models
if the representation of the symmetry group $G < Y\times Y$ is chosen appropriately.
The key part of the paper is section 4, where we confirm the above
 expectation for three infinite families of cases.
In section 5 we summarize the results, further
discuss the comparison with \cite{Obl}
and formulate open questions.
There is also an appendix containing background material.

\section{Quantum Hamiltonian reduction under polar actions}
\setcounter{equation}{0}

We here collect general definitions and results that will be used subsequently.
Our main purpose is to explain that formula (\ref{2.14}) characterizes
the reductions of the Laplace operator of a Riemannian manifold under
so-called polar actions  \cite{PT} of compact symmetry groups.
The exposition is restricted to
the necessary  minimum,
for more details see \cite{TMP} and references therein.

Let $Y$ be a smooth, connected, complete Riemannian
manifold with metric $\eta$.
Consider the Laplace operator $\Delta_Y$ corresponding to  $\eta$.
For a smooth function $F$,
in local coordinates $\{y^\mu\}$ on $Y$ one has
$\Delta_Y F = \vert \eta \vert^{-\frac{1}{2}}
\partial_\mu (\vert \eta\vert^{\frac{1}{2}} \partial^\mu F)$
with $\vert \eta \vert := \det (\eta_{\mu,\nu})$.
The restriction of  $\Delta_Y$ onto the space of
the complex-valued compactly supported smooth functions,
\be
\Delta_Y^0 := \Delta_Y|_{C_c^\infty(Y)}
\colon C_c^\infty(Y) \rightarrow C_c^\infty(Y),
\label{2.1}\ee
is an essentially self-adjoint linear operator
of the Hilbert space $L^2(Y, {\mathrm{d}} \mu_Y)$, where $\mu_Y$ denotes the
  measure generated by the Riemannian volume form,
  locally defined by $\vert \eta\vert^{\frac{1}{2}} \prod_\mu dy^\mu$.
 Suppose that a
\emph{compact} Lie group $G$ acts on $(Y, \eta)$ by isometries.
The action is given by a smooth map
\begin{equation}
\phi \colon G \times Y \rightarrow Y,
\quad
(g, y) \mapsto \phi(g, y) = \phi_g(y) = g . y
\label{2.2}\end{equation}
 such that
$\phi_g^* \eta = \eta$ for every $g \in G$.
The measure $\mu_Y$ inherits the  $G$-invariance and therefore
the Hilbert space $L^2(Y, {\mathrm{d}} \mu_Y)$ naturally carries a continuous unitary
representation of $G$.
This in turn is unitarily equivalent to an orthogonal  direct sum,
$L^2(Y, {\mathrm{d}} \mu_Y) \cong \oplus_\rho M_\rho
\otimes V_{\bar \rho}$, where $(\rho, V_\rho)$ runs over a
complete set of pairwise inequivalent
irreducible unitary representations of $G$,
$\bar \rho$ denotes the contragredient of the representation
$\rho$, and $M_\rho$ is a `multiplicity space' on which $G$ acts
trivially. Correspondingly, the self-adjoint scalar
Laplace  operator, $\bar \Delta_Y^0$, which by definition is the
closure of $\Delta_Y^0$  (\ref{2.1}), can be decomposed as
$\bar \Delta_Y^0 \cong \oplus_{\rho} \hat\Delta_\rho \otimes
{\mathrm{id}}_{V_{\bar \rho}}$, where $\hat \Delta_{\rho}$ is a
self-adjoint operator on the Hilbert space $M_{\rho}$.
The system $(M_\rho, \hat \Delta_{\rho})$ is  called the reduction of
the system $(L^2(Y, {\mathrm{d}} \mu_Y),\bar \Delta_Y^0)$
having the symmetry type $\bar \rho$.

In order to present a convenient model of $(M_\rho, \hat \Delta_{\rho})$,
consider now an irreducible unitary
representation
$(\rho, V)$ of $G$, where $V$ is a finite dimensional complex vector space with inner product
$(\, , )_V$. By simply acting componentwise, the differential operator $\Delta_Y^0$
extends onto the complex vector space of the $V$-valued compactly
supported smooth functions, $C_c^\infty(Y, V)$.
This gives the essentially self-adjoint operator
\begin{equation}
\Delta_Y^0 \colon C_c^\infty(Y, V) \rightarrow C_c^\infty(Y, V)
\label{2.3}\end{equation}
of the Hilbert space $L^2(Y, V, {\mathrm{d}} \mu_Y)$.
Because of the $G$-symmetry of the metric  $\eta$,
the set
\begin{equation}
C_c^\infty(Y, V)^G := \{ F \, | \, F \in C_c^\infty(Y, V),\,\,
F \circ \phi_g = \rho(g) \circ F \quad (\forall g \in G)\}
\label{2.4}\end{equation}
of the $V$-valued, compactly supported \emph{$G$-equivariant} smooth
functions is an invariant linear subspace of
$\Delta_Y^0$.
Moreover, the restriction of
$\Delta_Y^0$ (\ref{2.3})
onto $C_c^\infty(Y, V)^G$,
\begin{equation}
\Delta_\rho := \Delta_Y^0|_{C_c^\infty(Y, V)^G}
\colon C_c^\infty(Y, V)^G \rightarrow C_c^\infty(Y, V)^G,
\label{2.5}\end{equation}
is a densely defined, symmetric, essentially self-adjoint linear operator on
the Hilbert space
$L^2(Y, V, {\mathrm{d}} \mu_Y)^G$
of the  square-integrable $G$-equivariant functions.
It is not
difficult to demonstrate the unitary equivalence
\begin{equation}
(M_\rho, \hat \Delta_\rho)\cong (L^2(Y,V, {\mathrm{d}}\mu_Y)^G, \bar
\Delta_\rho) \quad\hbox{with}\qquad V:= V_\rho,
\label{redsyst}\end{equation} where $\bar \Delta_\rho$ denotes the
closure of $\Delta_\rho$ in (\ref{2.5}).
It is convenient for many purposes
to use the realization of the reduced quantum system
furnished by $L^2(Y,V, {\mathrm{d}}\mu_Y)^G$.

Particularly simple cases of the reduction arise
if the reduced configuration space
$Y_{\mathrm{red}} := Y / G$ is a smooth manifold,
although this happens very rarely.
However, restricting to the principal orbit type,
$\check{Y} \subset Y$, one always obtains a smooth fiber bundle
$\pi \colon \check{Y} \rightarrow \check{Y} / G$.
Note that $\check Y$  consists of the points of $Y$ having the
smallest isotropy subgroups for the $G$-action \cite{GOV}.
The `big cell'  of the reduced configuration space,
given by $\check{Y}_{\mathrm{red}} := \check{Y} / G$,
is naturally endowed with a
Riemannian metric, $\eta_\mathrm{red}$, making $\pi$ a Riemannian
submersion. From a quantum mechanical point of view, neglecting the
non-principal orbits is harmless, in some sense, since $\check{Y}$
is not
only open and dense in $Y$, but it is also of full
measure.

In many applications  \emph{polar group actions} are important, whose characteristic property
is that the $G$-orbits  possess representatives that form
\emph{sections} in the sense of Palais and Terng \cite{PT}.
By definition, a section $\Sigma \subset Y$ is a connected, closed, regularly
embedded smooth submanifold of $Y$ that meets every $G$-orbit and it does
so orthogonally at every intersection point of $\Sigma$ with an
orbit.
If a section exists, then any two sections are $G$-related.
The induced metric on $\Sigma$ is denoted by $\eta_\Sigma$, and
for the  measure generated by $\eta_\Sigma$ we introduce the
notation $\mu_\Sigma$.
For a section $\Sigma$, denote by $\check{\Sigma}$ a
connected component of the manifold
$\hat{\Sigma} := \check{Y} \cap \Sigma$.
The isotropy subgroups of all elements of $\hat{\Sigma}$ are the same
and for a fixed section we define $K := G_y$ for $y \in \hat{\Sigma}$.
The group $K$ is called the centralizer of the section $\Sigma$.
By restricting $\pi \colon \check{Y} \rightarrow \check{Y} / G$
onto $\check{\Sigma}$,
$(\check{Y}_{\mathrm{red}}, \eta_\mathrm{red})$ becomes identified
with $(\check{\Sigma}, \eta_{\check{\Sigma}})$, where $\eta_{\check{\Sigma}}$
is the induced  metric on $\check{\Sigma}$.
We let $\Delta_{\check \Sigma}$ stand for
the Laplace  operator of the Riemannian manifold
$(\check \Sigma, \eta_{\check \Sigma})$.
The $G$-equivariant diffeomorphism
\begin{equation}
\check{\Sigma} \times (G / K) \ni (Q, g K) \mapsto \phi_g(Q) \in \check{Y}
\label{2.7}\end{equation}
provides a trivialization of the fiber bundle
$\pi \colon \check{Y} \rightarrow \check{Y} / G$.
Generalized polar coordinates on $\check Y$
consist of `radial' coordinates
on $\check \Sigma$ and  `angular' coordinates on $G/K$.

To concretize
the reduced system (\ref{redsyst}) for polar actions,
we introduce the  space
\begin{equation}
\mathrm{Fun}(\check{\Sigma}, V^K)
:= \{ f \, | \, f \in C_c^\infty(\check{\Sigma}, V^K),
f = F|_{\check{\Sigma}} \mbox{ for some }
F \in C_c^\infty(Y, V)^G \},
\label{2.8}\end{equation}
where $V^K$ is spanned by the $K$-invariant vectors in the
representation space $V$. We assume that the representation $(\rho, V)$ of
the symmetry group $G$ is admissible
in the sense that
\be
\dim(V^K) > 0.
\label{2.9}\ee
The restriction of functions appearing in the definition (\ref{2.8})
gives
rise to
a linear isomorphism
$\mathrm{Fun}(\check{\Sigma}, V^K) \cong C_c^\infty(Y, V)^G \hookrightarrow
L^2(Y, V, {\mathrm{d}} \mu_Y)^G$.
This induces
a scalar product on $\mathrm{Fun}(\check{\Sigma}, V^K)$  making it a
pre-Hilbert space whose closure satisfies the Hilbert space
isomorphism
$\mathrm{\overline{Fun}}(\check{\Sigma}, V^K) \cong
L^2(Y, V, {\mathrm{d}} \mu_Y)^G$.
Next,  consider the Lie algebra ${\mathcal{G}} := \mathrm{Lie}(G)$ and its
subalgebra
${\mathcal{K}}:= \mathrm{Lie}(K)$.
Fix a $G$-invariant positive definite
scalar product, ${\mathcal{B}}_\cG$, on ${\mathcal{G}}$  and thereby
determine the orthogonal complement ${\mathcal{K}}^\perp$ of
${\mathcal{K}}$ in $\mathcal{G}$.
For any $\xi \in {\mathcal{G}}$
denote by $\xi^{\sharp}$ the associated vector field on $Y$.
Then at each point
$\mQ \in \check{\Sigma}$ the linear map
${\mathcal{K}}^\perp \ni \xi \mapsto \xi^{\sharp}_\mQ \in T_\mQ Y$
is injective, and the \emph{inertia operator}
$J(\mQ) \in \mathrm{End}({\mathcal{K}}^\perp)$ can
be defined by the requirement
\begin{equation}
\eta_\mQ(\xi^\sharp_\mQ, \zeta^\sharp_\mQ) =
{\mathcal{B}}_\cG(\xi, J(\mQ) \zeta),
\qquad
\forall \xi, \zeta \in {\mathcal{K}}^\perp.
\label{J def}
\end{equation}
Note that $J(\mQ)$ is symmetric and positive definite with
respect to
${\mathcal{B}}_\cG\vert_{{\mathcal{K}}^\perp\times {\mathcal{K}}^\perp}$.
By choosing dual
bases $\{ T_\alpha \}$, $\{ T^\alpha \}$
$\subset {\mathcal{K}}^\perp$, that is,
${\mathcal{B}}_\cG(T^\alpha, T_\beta) = \delta^\alpha_\beta$,
we let
\begin{equation}
b_{\alpha, \beta}(\mQ) := {\mathcal{B}}_\cG(T_\alpha, J(\mQ) T_\beta),
\qquad
b^{\alpha, \beta}(\mQ) := {\mathcal{B}}_\cG(T^\alpha, J(\mQ)^{-1} T^\beta).
\label{b-def}\end{equation}

The $G$-orbit $G . \mQ \subset Y$ through any point $\mQ \in \check{\Sigma}$ is
an embedded submanifold of $Y$ and by its embedding it inherits
a Riemannian metric, $\eta_{G . \mQ}$.
Thus we can define  the smooth \emph{density function}
$\delta \colon \check{\Sigma} \rightarrow (0, \infty)$ by
\begin{equation}
\delta(\mQ) := \mbox{volume of the Riemannian manifold } (G . \mQ, \eta_{G . \mQ}),
\label{2.12}\end{equation}
where the volume is understood with respect to
the measure, $\mu_{G . \mQ}$,  belonging to $\eta_{G . \mQ}$.
It is easy to see that
\be
\delta(\mQ) = C\vert \det (J(\mQ))\vert^{\frac{1}{2}} =
C  \vert \det \left(b_{\alpha, \beta}(\mQ)\right)\vert^{\frac{1}{2}}
\label{2.13}\ee
with some constant $C>0$.
In the following proposition, quoted from \cite{TMP},
$\rho\prime$   denotes the representation of $\mathcal{G}$ corresponding to the
representation $\rho$ of $G$.

\medskip
\noindent
\textbf{Proposition 2.1 }
\emph{Let us consider a polar $G$-action using the above notations.
Then the reduced system (\ref{redsyst}) associated with
an admissible irreducible  unitary representation $(\rho, V)$ of
$G$ can be identified with the pair
$(L^2(\check{\Sigma}, V^K, {\mathrm{d}} \mu_{\check{\Sigma}}),
\Delta_{\mathrm{red}})$, where
\begin{equation}
\Delta_{\mathrm{red}}
= \Delta_{\check{\Sigma}} - \delta^{-\frac{1}{2}}
\Delta_{\check{\Sigma}}(\delta^{\frac{1}{2}})
+ b^{\alpha, \beta} \rho\prime(T_\alpha) \rho\prime(T_\beta)
\label{2.14}\end{equation}
with domain
${\mathcal{D}}(\Delta_{\mathrm{red}})
= \delta^{\frac{1}{2}} \mathrm{Fun}(\check{\Sigma}, V^K)$
is a densely defined, symmetric, essentially self-adjoint operator on
the Hilbert space
$L^2(\check{\Sigma}, V^K, {\mathrm{d}} \mu_{\check{\Sigma}})$.}
\medskip

The above statement results by calculating the action of $\Delta_Y$ on the $V$-valued equivariant
functions in (\ref{2.8})
with the aid of polar coordinates,  using also
the Hilbert space identifications
\begin{equation}
\overline{\mathrm{Fun}}(\check \Sigma, V^K) \cong
L^2(Y, V, {\mathrm{d}} \mu_Y)^G
\cong L^2(\check \Sigma, V^K, \delta
{\mathrm{d}}\mu_{\check \Sigma}).
\label{2.15}\end{equation}
The last equality follows  by integrating out the `angular' coordinates
in the scalar product of equivariant functions.
One also uses the unitary map $U: L^2(\check \Sigma, V^K, \delta \mathrm{d}\mu_{\check \Sigma})\to
L^2(\check \Sigma, V^K,  \mathrm{d}\mu_{\check \Sigma})$ defined by
$U: f \mapsto \delta^{\frac{1}{2}} f$.

The first term in (\ref{2.14})
corresponds to the kinetic energy of a particle moving
on $(\check Y_{\mathrm{red}}, \eta_{\mathrm{red}})\cong
(\check \Sigma, \eta_{\check \Sigma})$  and the rest
represents potential energy if
$\mathrm{dim}(V^K)=1$.
The second term of (\ref{2.14}) is always  potential
energy, which is constant in some cases. We refer to this term as the `measure factor'.
It represents a significant difference between the outcomes of the corresponding classical and
quantum Hamiltonian reductions \cite{TMP}.
If $\mathrm{dim}(V^K)>1$,
then one says that the reduced system contains internal `spin'
degrees of freedom and then
the third term of (\ref{2.14}) encodes `spin-dependent potential energy'.

\section{Examples of polar actions on compact Lie groups}
\setcounter{equation}{0}

From now we take the `unreduced configuration space'  $Y$ to be  a compact,
connected,  real Lie group endowed with a bi-invariant
metric $\eta$, induced by a positive definite, $Y$-invariant
bilinear form $\cB_\cY$ of the Lie algebra $\cY := \Lie(Y)$.
For the reduction group $G$ one may choose any symmetric subgroup
of the direct product group $Y \times Y$, that is,
\be
(Y \times Y)^\sigma_0 \leq G \leq (Y \times Y)^\sigma,
\label{3.1}\ee
where $(Y \times Y)^\sigma$ stands for the fixed-point set of some
involutive  automorphism $\sigma \in \Inv(Y \times Y)$, and
$(Y \times Y)^\sigma_0$
is the connected component of the identity in $(Y \times Y)^\sigma$.
The group $G$ acts on $Y$ by the map
\be
\phi \colon G \times Y \rightarrow Y, \quad
((g_L, g_R), y) \mapsto \phi_{(g_L, g_R)}(y)
:= g_L y g_R^{-1}.
\label{phi}\ee
The group actions of this form are often called Hermann actions.
Under mild conditions, which hold in the examples below,
these  are polar actions in the sense of \cite{PT}.
In fact, the sections are provided by certain toral subgroups\footnote{A toral subgroup $A<Y$
is a connected and closed Abelian subgroup.
It is the closedness of the relevant subgroups that requires some conditions.
If  $\cY$ is semi-simple, then a sufficient condition is to take $\cB_\cY$ as a multiple of the Killing form
\cite{HPTT}.}
$A < Y$.
Thus the sections are flat in the induced metric, which is the characteristic
property of the so-called hyperpolar actions \cite{HPTT}.
In the simplest  special case
$\sigma(y_1, y_2) = (y_2, y_1)$,
$G = Y_\diag = \{ (y, y) \, | \, y \in Y \} \cong Y$ and
(\ref{phi}) is just the adjoint action
of $Y$ on itself, for which the sections are the maximal tori of $Y$.

\subsection{Hermann actions associated with pairs of involutions}

The reductions that we study later arise from the following construction.
Let $\sigma_L, \sigma_R \in \Inv(Y)$ be two
involutions of $Y$, and let $Y_L, Y_R \leq Y$ be corresponding
symmetric subgroups of $Y$,
\be
(Y^{\sigma_I})_0 \leq Y_I \leq Y^{\sigma_I}
\quad
(I \in \{L, R\}).
\label{3.3}\ee
We suppose that the scalar product $\cB_\cY$ is invariant under both
$\sigma_L$ and $\sigma_R$ and introduce
$\sigma \in \Inv(Y\times Y)$ by
$\sigma(y_1, y_2) := (\sigma_L(y_1), \sigma_R(y_2))$.
Then
\be
G := Y_L \times Y_R
\ee
is a symmetric subgroup of $Y\times Y$ and
equation (\ref{phi}) defines a  hyperpolar Hermann action of $G$ on $Y$.
The classification of the inequivalent pairs of involutions $(\sigma_L, \sigma_R)$ has been
worked out by Matsuki \cite{Mats_02}.
We assume for simplicity that the two involutions $\sigma_L$ and $\sigma_R$
commute with each other, which holds for the large majority of  cases in the classification.
Subsequently, the induced Lie
algebra involutions  are denoted by the same letters $\sigma_L$ and $\sigma_R$.

Now, with the aid of the subspaces
\be
\cY^{\sigma_I, \pm} := \ker(\sigma_I \mp \Id_\cY) \subset \cY
\quad
(I \in \{ L, R \})
\quad
\mathrm{and}
\quad
\cY^{\pm \pm} := \cY^{\sigma_L, \pm} \cap \cY^{\sigma_R, \pm} \subset \cY
\label{3.4}\ee
we obtain the orthogonal decomposition
\be
\cY = \cY^{++} \oplus \cY^{+-} \oplus \cY^{-+} \oplus \cY^{--},
\label{gradation}
\ee
which gives also a $\bZ_2 \times \bZ_2$-gradation of $\cY$.
The Lie algebra of the symmetric subgroup $Y_I \leq Y$
is $\Lie(Y_I) \cong \cY^{\sigma_I, +}$ $(I \in \{ L, R \})$.
Then, we choose a \emph{maximal
Abelian subalgebra} $\cA$ in $\cY^{--}$  and also define $A := \exp(\cA)$,
which is a toral subgroup of $Y$.
According to an important theorem proved in \cite{Hoog, Mats_97},
the Lie group $Y$ admits the generalized Cartan decomposition
\be
Y = Y_L A Y_R.
\label{3.6}\ee
This means that every element of $Y$ can be written as a product of the
elements of the subgroups in (\ref{3.6}).
Recalling the definition of the
Hermann action (\ref{phi}) for $G = Y_L \times Y_R$, equation (\ref{3.6}) says
that the subgroup $A$ intersects every $G$-orbit. Moreover,
it does so orthogonally at every intersection point, and thus $A$ provides
a \emph{section} for the $G$-action in the sense of \cite{PT}.
Below $\check A$ denotes a connected component of the regular part of the
section $A$.

Let us introduce the subgroups $Y_{L R} := Y_L \cap Y_R \leq Y$ and
\be
M :=  \{ g \, | \, g \in Y_{L R}, \,\, g a g^{-1}=a
\,\, (\forall a \in A) \}
\leq Y_{L R}.
\label{M_def}
\ee
Their Lie algebras are
\begin{align}
& \Lie(Y_{L R}) \cong \Lie(Y_L) \cap \Lie(Y_R)
\cong \cY^{\sigma_L, +} \cap \cY^{\sigma_R, +} = \cY^{+ +},
\label{3.8}\\
& \cM := \Lie(M)
= \{ X \, | \, X \in \cY^{+ +}, \,\, \ad_X(\bfq) = 0
\,\,(\forall \bfq \in \cA) \},
\label{3.9}\end{align}
where $\ad_X$ is defined by the Lie bracket on $\cY$.
It can be shown that the centralizer of the section $A = \exp(\cA)$
(the isotropy subgroup of the elements of $\check A$)
is now furnished by
\be
K = M_{\diag} = \{ (g, g) \, | \, g \in M \} \leq G.
\label{3.10}\ee

To specialize the inertia operator $J$ defined in (\ref{J def}),
we introduce a $G$-invariant scalar product on the Lie algebra
\be
\cG = \Lie(G) = \Lie(Y_L \times Y_R)
\cong \Lie(Y_L) \oplus \Lie(Y_R)
\cong \cY^{\sigma_L, +} \oplus \cY^{\sigma_R, +}
\label{3.11}\ee
by the formula
\be
\cB_\cG( (\xi_L, \xi_R), (\zeta_L, \zeta_R) )
:= \cB_\cY(\xi_L, \zeta_L) + \cB_\cY(\xi_R, \zeta_R),
\qquad
\forall (\xi_L, \xi_R), (\zeta_L, \zeta_R) \in \cG.
\label{3.12}\ee
This induces the decomposition  $\cG = \cK \oplus \cK^\perp$, where $\cK = \Lie(K)$.
By using the decomposition $\cY = \cM \oplus \cM^\perp$ defined by $\cB_\cY$,
we also  introduce the subspaces
\begin{align}
& \cK^\perp_a := \{ (X, -X) \, | \, X \in \cM \} \subset \cK^\perp,
\label{M_-} \\
& \cK^\perp_e := \{ (\xi_L, \xi_R) \, | \, \xi_L, \xi_R \in \cM^\perp \cap \cY^{++} \}
\subset \cK^\perp,
\label{S_e} \\
& \cK^\perp_o := \{ (\zeta_L, \zeta_R) \, | \, \zeta_L \in \cY^{+-},
\zeta_R \in  \cY^{-+} \} \subset \cK^\perp,
\label{S_o}
\end{align}
which yield the orthogonal decomposition
\be
\cK^\perp = \cK^\perp_a \oplus \cK^\perp_e \oplus \cK^\perp_o.
\label{3.16}\ee
Now consider the vector field
$\xi^\sharp = (\xi_L, \xi_R)^\sharp$
 on $Y$ associated with  $\xi = (\xi_L, \xi_R) \in \cG$ by means of the  $G$-action.
 At an arbitrary
point $e^\bfq \in A$ $(\bfq \in \cA)$ of the section $A$ we find
\be
\xi^\sharp_{e^\bfq} = (\xi_L, \xi_R)^\sharp_{e^\bfq}
= (\dd L_{e^\bfq})_e \left( \xi_R - e^{-\ad_\bfq}(\xi_L) \right)
\in T_{e^\bfq} Y,
\label{xiSharp}
\ee
where $L_y$ denotes the left-translation on $Y$
by group element $y \in Y$.
Simply by plugging (\ref{xiSharp}) into the definition (\ref{J def}),
routine algebraic manipulations lead to the following result.

\begin{lemma} Equation (\ref{3.16}) is a decomposition of $\cK^\perp$
into invariant subspaces
of the inertia operator $J(e^\bfq)$ at any point $e^\bfq \in \check{A}$.
One has $J(e^\bfq)\big{|}_{\cK^\perp_a} = 2 \Id_{\cK^\perp_a}$
and, writing $\xi = (\xi_L, \xi_R) \in \cG$ as a
$2$-component column vector with components $\xi_L$ and $\xi_R$,  the action
of  $J(e^\bfq)$ on $\cK^\perp_e$ and $\cK^\perp_o$ is encoded by the matrices
\be
J(e^\bfq)\big{|}_{\cK^\perp_e} = \begin{bmatrix}
\1 & -\cosh(\ad_\bfq) \\
-\cosh(\ad_\bfq) & \1
\end{bmatrix} \bigg{|}_{\cK^\perp_e},
\quad
J(e^\bfq)\big{|}_{\cK^\perp_o} = \begin{bmatrix}
\1 & -\sinh(\ad_\bfq) \\
\sinh(\ad_\bfq) & \1
\end{bmatrix} \bigg{|}_{\cK^\perp_o}.
\label{J_on_S}
\ee
For the inverse of $J(e^\bfq)$ one has
$J(e^\bfq)^{-1}\big{|}_{\cK^\perp_a} = \frac{1}{2} \Id_{\cK^\perp_a}$
together with
\begin{align}
& J(e^\bfq)^{-1}\big{|}_{\cK^\perp_e} = - \begin{bmatrix}
\sinh^{-2}(\ad_\bfq) & \cosh(\ad_\bfq) \sinh^{-2}(\ad_\bfq) \\
\cosh(\ad_\bfq) \sinh^{-2}(\ad_\bfq) & \sinh^{-2}(\ad_\bfq)
\end{bmatrix} \bigg{|}_{\cK^\perp_e}, &
\label{3.19}\\
& J(e^\bfq)^{-1}\big{|}_{\cK^\perp_o} = \begin{bmatrix}
\cosh^{-2}(\ad_\bfq) & \sinh(\ad_\bfq) \cosh^{-2}(\ad_\bfq) \\
-\sinh(\ad_\bfq) \cosh^{-2}(\ad_\bfq) & \cosh^{-2}(\ad_\bfq)
\end{bmatrix} \bigg{|}_{\cK^\perp_o}. &
\label{3.20}\end{align}
\end{lemma}

\subsection{A family of two involutions on $U(N)$}

For our later purpose we now focus on the unitary group
\be
Y := U(N) = \{ y \, | \, y \in GL(N, \bC), \quad y^\dagger y = \1_N \}.
\label{3.21}\ee
We equip the Lie algebra
\be
\cY := \mfu(N) = \{ X \, | \, X \in \mfgl(N, \bC), \quad X^\dagger + X = 0 \}
\label{3.22}\ee
with the scalar product
\be
\cB_\cY(X, Z) := - \tr(X Z),
\qquad
\forall X, Z \in \mfu(N).
\label{3.23}\ee
To any pair $(m, n) \in \bZ_{\geq 0}^2$ with $m \geq n$ and $m + n = N$ we
associate the  block-matrix
\be
\bfI_{m, n} := \diag(\1_m, -\1_n)
= \begin{bmatrix}
\1_m & 0 \\
0 & -\1_n
\end{bmatrix} \in U(N),
\label{3.24}\ee
and the involutive inner automorphism
\be
\theta_{m, n} \colon U(N) \rightarrow U(N),
\quad
y \mapsto \theta_{m, n}(y) := \bfI_{m, n} y \bfI_{m, n}^{-1}.
\label{3.25}\ee
The fixed-point set of $\theta_{m, n}$ is
\be
U(N)^{\theta_{m, n}}
= \left \{
\begin{bmatrix}
a & 0 \\
0 & b
\end{bmatrix}
\, \bigg{|} \, a \in U(m), b \in U(n) \right \}
\cong U(m) \times U(n).
\label{3.26}\ee
Note that $U(N)^{\theta_{m, n}}$ is connected.
The induced Lie algebra involution operates as
\be
\theta_{m, n}(X) = \bfI_{m, n} X \bfI_{m, n}^{-1},
\qquad
\forall X \in \mfu(N).
\label{3.27}\ee
Using the block-matrix realization
\be
\mfu(N) = \left \{
\begin{bmatrix}
A & C \\
- C^\dagger & B
\end{bmatrix}
\, \bigg{|} \,
A \in \mfu(m), B \in \mfu(n), C \in \bC^{m \times n} \right \},
\label{3.28}\ee
the eigenspaces
$\mfu(N)^{\theta_{m, n}, \pm}$ are
\begin{align}
\mfu(N)^{\theta_{m, n}, +} = \left \{
\begin{bmatrix}
A & 0 \\
0 & B
\end{bmatrix}
\, \bigg{|} \,
A \in \mfu(m), B \in \mfu(n) \right \},
\,\,
\mfu(N)^{\theta_{m, n}, -} = \left \{
\begin{bmatrix}
0 & C \\
- C^\dagger & 0
\end{bmatrix}
\, \bigg{|} \,
C \in \bC^{m \times n} \right \}.
\label{mfu_N_subspaces}
\end{align}

Now we take two pairs $(m, n), (r, s) \in \bZ_{\geq 0}^2$ with the additional
requirements $m \geq r \geq s \geq n$ and $m + n = r + s = N$, and
consider the commuting involutions
\be
\sigma_L := \theta_{r, s}
\quad \mbox{and} \quad
\sigma_R := \theta_{m, n}.
\label{3.30}\ee
The corresponding symmetric subgroups $Y_L, Y_R \leq Y$ are
\be
U(N)_L := U(N)^{\sigma_L} \cong U(r) \times U(s)
\quad \mbox{and} \quad
U(N)_R := U(N)^{\sigma_R} \cong U(m) \times U(n).
\label{3.31}\ee
The partition $N = n + (r - n) + (s - n) + n$ leads to a $4 \times 4$
block-matrix
decomposition of any $N \times N$ matrix in general.
(Of course, if $r = n$ or $s = n$, then the  block-matrix
decomposition contains fewer blocks.)
 That is,
any matrix $X \in \bC^{N \times N}$ can be written as
\be
X = \begin{bmatrix}
X_{1, 1} & X_{1, 2} & X_{1, 3} & X_{1, 4} \\
X_{2, 1} & X_{2, 2} & X_{2, 3} & X_{2, 4} \\
X_{3, 1} & X_{3, 2} & X_{3, 3} & X_{3, 4} \\
X_{4, 1} & X_{4, 2} & X_{4, 3} & X_{4, 4}
\end{bmatrix},
\label{matrixDecomp}
\ee
where the entries $X_{i, j}$ are themselves matrices,
$X_{1, 1} \in \bC^{n \times n}$, $X_{1, 2} \in \bC^{n \times (r - n)}$,
$X_{1, 3} \in \bC^{n \times (s - n)}$, $X_{1, 4} \in \bC^{n \times n}$, etc.
Then for the Lie group $Y_{LR} = Y_L \cap Y_R$ we have
\be
U(N)_{LR}
= \left \{
\begin{bmatrix}
a_{1, 1} & a_{1, 2} & 0 & 0 \\
a_{2, 1} & a_{2, 2} & 0 & 0 \\
0 & 0 & a_{3, 3} & 0 \\
0 & 0 & 0 & a_{4, 4}
\end{bmatrix}
\, \Bigg{|} \,
\begin{bmatrix}
a_{1, 1} & a_{1, 2} \\
a_{2, 1} & a_{2, 2}
\end{bmatrix}
\in U(r),
a_{3, 3} \in U(s - n), a_{4, 4} \in U(n)
\right \}.
\label{Y_LR}
\ee
Therefore
$U(N)_{LR} \cong U(r) \times U(s - n) \times U(n)$ and
the Lie algebra
$\Lie(U(N)_{LR}) = \mfu(N)^{+ +}$
is isomorphic to
$ \mfu(r) \oplus \mfu(s - n) \oplus \mfu(n)$.
In our case the subspace $\cY^{--}$ in (\ref{3.4})
reads
\be
\mfu(N)^{--} = \left \{
\begin{bmatrix}
0 & 0 & 0 & A_{1, 4} \\
0 & 0 & 0 & A_{2, 4} \\
0 & 0 & 0 & 0 \\
- A_{1, 4}^\dagger & - A_{2, 4}^\dagger & 0 & 0
\end{bmatrix}
\, \Bigg{|} \,
A_{1, 4} \in \bC^{n \times n}, A_{2, 4} \in \bC^{(r - n) \times n}
\right \}.
\label{3.34}\ee
To proceed, we define
the diagonal $n \times n$ matrix
\be
q := \diag(q_1, q_2, \ldots, q_n) \in \bR^{n \times n}
\label{q}\ee
for  any real $n$-tuple $(q_1, q_2,\ldots , q_n) \in \bR^n$,
and we also set
\be
\bfq := \begin{bmatrix}
0 & 0 & 0 & q \\
0 & 0 & 0 & 0 \\
0 & 0 & 0 & 0 \\
-q & 0 & 0 & 0
\end{bmatrix}
\in \mfu(N)^{--}.
\label{bfq}\ee
Then  the set of matrices
\be
\cA := \{ \bfq \, | \, (q_1, q_2,\ldots , q_n) \in \bR^n \} \subset \mfu(N)^{--}
\label{3.37}\ee
is a maximal Abelian subalgebra in $\mfu(N)^{--}$.
A basis of the  dual space $\cA^*$ is given by the functionals
\be
\epsilon_k \colon \cA \rightarrow \bR,
\quad
\bfq \mapsto \epsilon_k(\bfq) := q_k.
\label{epsilon_k}
\ee
The corresponding subgroup $A = \exp(\cA)$
has the form
\be
A = \left \{ e^\bfq =  \begin{bmatrix}
\cos(q) & 0 & 0 & \sin(q) \\
0 & \1_{r - n} & 0 & 0 \\
0 & 0 & \1_{s - n} & 0 \\
-\sin(q) & 0 & 0 & \cos(q)
\end{bmatrix}
\, \Bigg{|} \,
(q_1, q_2,\ldots , q_n) \in \bR^n
\right \}.
\label{form_of_A}
\ee
If $\bT(n)$ denotes
the diagonally embedded standard torus in $U(n)$,
then it is straightforward to show that
the subgroup $M$ (\ref{M_def}) is now furnished by
\be
M = \left \{ \begin{bmatrix}
a & 0 & 0 & 0 \\
0 & b & 0 & 0 \\
0 & 0 & c & 0 \\
0 & 0 & 0 & a
\end{bmatrix}
\, \Bigg{|} \,
a \in \bT(n),\, b \in U(r - n),\, c \in U(s - n)
\right \}.
\label{M}\ee
Note that $M$ is connected, and therefore so is
the centralizer $K = M_\diag$ of the section $A$.
Moreover, we  have the
identifications
\be
K \cong M_\diag \cong
M \cong \bT(n) \times U(r - n) \times U(s - n)
\cong U(1)^{\times n} \times U(r - n) \times U(s - n).
\label{3.41}\ee

It is shown in \cite{Mats_97} (page 63) that the closed, connected subset
\be
A_+ := \left \{ e^\bfq
\, \Big{|} \,
0 \leq q_1 \leq q_2 \leq \ldots \leq q_n \leq \frac{\pi}{2} \right \}
\subset A
\label{3.42}\ee
intersects each orbit of
$G=U(N)^{\sigma_L} \times U(N)^{\sigma_R}$ under the action (\ref{phi})
precisely once.
Note also that matrix exponentiation provides a bijection from
\be
\cA_+ := \left \{ \bfq
\, \Big{|} \,
0 \leq q_1 \leq q_2 \leq \ldots \leq q_n \leq \frac{\pi}{2} \right \}
\subset \cA
\label{3.43}\ee
onto $A_+$.
By inspecting the isotropy subgroup $G_{e^\bfq} \leq G$  for
$e^\bfq \in A_+$,  we find that
$G_{e^\bfq} = K$
if and only if
$\bfq \in \check{\cA}_+$, where $\check{\cA}_+$ denotes
the connected open subset
\be
\check{\cA}_+ := \left \{ \bfq
\, \Big{|} \,
0 < q_1 < q_2 < \ldots < q_n < \frac{\pi}{2} \right \}
\subset \cA_+.
\label{cA_check_+}
\ee
We can conclude from the above that
the subset $\check{A} := \exp(\check{\cA}_+)$ provides a connected
component for the regular part of the section $A$.
Regarding the components $q_k$ in (\ref{cA_check_+}) as global
coordinates on $\check{A}$, for the Laplace operator
$\Delta_{\check{A}}$ defined by the induced metric
we obtain
\be
\Delta_{\check{A}} = \frac{1}{2} \sum_{k = 1}^n \frac{\partial^2}{\partial q_k^2}.
\label{LB_on_checkA}
\ee

\subsection{Diagonalization of the inertia operator}

We continue the study of the examples (\ref{3.30}) by presenting
a basis of $\cK^\perp$ that diagonalizes $J(e^\bfq)$ (\ref{J_on_S}) for any $\bfq\in \check{\cA}_+$
in (\ref{cA_check_+}).
We then use this basis to compute the density $\delta^{\frac{1}{2}}$ that enters
the second term of the reduced Laplacian (\ref{2.14}).
Note that  $\delta^{\frac{1}{2}}$  could  be found also
by the specialization of general formulae
available for two commuting involutions  \cite{Hoog,Heck}, but
we need to fix a basis for the evaluation of the third term of (\ref{2.14}),
which will be performed later.

We start by defining an orthonormal basis (ONB) in the space
$\cM^\perp \cap \mfu(N)^{+ +}$, which (due to (\ref{Y_LR}) and (\ref{M}))  has the form
\be
\cM^\perp \cap \mfu(N)^{+ +} = \left\{
\begin{bmatrix}
X_{1, 1} & X_{1, 2} & 0 & 0 \\
-X_{1, 2}^\dagger & 0 & 0 & 0 \\
0 & 0 & 0 & 0 \\
0 & 0 & 0 & X_{4, 4}
\end{bmatrix}
\, \Bigg| \,
\begin{array}{l}
X_{1, 1}, X_{4, 4} \in \mfu(n),
(X_{1, 1} + X_{4, 4})_\diag = 0, \\
X_{1, 2} \in \bC^{n \times (r - n)}
\end{array}
\right\}.
\label{3.47}\ee
If $r = n$, then  there are no off-diagonal blocks, and in general
$\dim(\cM^\perp \cap \mfu(N)^{+ +}) = n (2 r - 1)$.
For all $1 \leq j \leq n$ we let
\be
E_{2 \epsilon_j}^\ri
:= \frac{\ri}{\sqrt{2}}
\begin{bmatrix}
E_{j j} & 0 & 0 & 0 \\
0 & 0 & 0 & 0 \\
0 & 0 & 0 & 0 \\
0 & 0 & 0 & -E_{j j}
\end{bmatrix},
\label{3.48}\ee
and for all $1 \leq k < l \leq n$ we define
$$
 E_{\epsilon_k + \epsilon_l}^\rr
:= \frac{1}{2}
\begin{bmatrix}
E_{k l} - E_{l k} & 0 & 0 & 0 \\
0 & 0 & 0 & 0 \\
0 & 0 & 0 & 0 \\
0 & 0 & 0 &  E_{l k} - E_{k l}
\end{bmatrix},\,\,
E_{\epsilon_k + \epsilon_l}^\ri
:= \frac{\ri}{2}
\begin{bmatrix}
E_{k l} + E_{l k} & 0 & 0 & 0 \\
0 & 0 & 0 & 0 \\
0 & 0 & 0 & 0 \\
0 & 0 & 0 & - E_{k l} - E_{l k}
\end{bmatrix},
$$
\be
 E_{\epsilon_k - \epsilon_l}^\rr
:= \frac{1}{2}
\begin{bmatrix}
E_{k l} - E_{l k} & 0 & 0 & 0 \\
0 & 0 & 0 & 0 \\
0 & 0 & 0 & 0 \\
0 & 0 & 0 & E_{k l} - E_{l k}
\end{bmatrix},\,\,
 E_{\epsilon_k - \epsilon_l}^\ri
:= \frac{\ri}{2}
\begin{bmatrix}
E_{k l} + E_{l k} & 0 & 0 & 0\\
0 & 0 & 0 & 0 \\
0 & 0 & 0 & 0 \\
0 & 0 & 0 & E_{k l} + E_{l k}
\end{bmatrix}.
\label{3.49}\ee
 For all $1 \leq j \leq n$ and
$1 \leq d \leq r - n$ we set
\begin{align}
& E_{\epsilon_j}^{\rr, d}
:=  \frac{1}{\sqrt{2}}
\begin{bmatrix}
0 & E_{j d} & 0 & 0 \\
-E_{d j} & 0 & 0 & 0 \\
0 & 0 & 0 & 0 \\
0 & 0 & 0 & 0
\end{bmatrix}, &
& E_{\epsilon_j}^{\ri, d}
:=  \frac{\ri}{\sqrt{2}}
\begin{bmatrix}
0 & E_{j d} & 0 & 0 \\
E_{d j} & 0 & 0 & 0\\
0 & 0 & 0 & 0 \\
0 & 0 & 0 & 0
\end{bmatrix}. &
\end{align}
The superscripts $\ri$ and $\rr$ refer  to purely
imaginary and to real matrices, respectively, and  the elementary matrices
$E_{ab}$ are always understood to be of the correct size as dictated by
(\ref{matrixDecomp}).
The set of matrices
\be
\{ E_\alpha^D \}_{\alpha, D} :=
\{ E_{\epsilon_k \pm \epsilon_l}^\rr, E_{\epsilon_k \pm \epsilon_l}^\ri\}_{1 \leq k < l \leq n}
\cup
\{ E_{2 \epsilon_j}^\ri \}_{j = 1}^n
\cup
\{ E_{\epsilon_j}^{\rr, d}, E_{\epsilon_j}^{\ri, d} \}_{\begin{subarray}{l}
1 \leq j \leq n, \\
1 \leq d \leq r - n
\end{subarray}}
\ee
forms an ONB in $\cM^\perp \cap \mfu(N)^{+ +}$. Here $D$ is an `index
of degeneration' and  $\alpha$ runs over the positive
roots $\cR_+$ for the root system $C_n$ or $BC_n$. More precisely,
\be
\cR_+ = \begin{cases}
\cR_+(C_n) & \text{if $r = n$}, \\
\cR_+(BC_n) & \text{if $r > n$}.
\end{cases}
\ee
One can easily verify the relations
\be
(\ad_\bfq)^2 E_\alpha^D = -\alpha(\bfq)^2 E_\alpha^D.
\label{commut1}\ee

Next, we deal with the subspaces $\mfu(N)^{+ - }$ and $\mfu(N)^{- +}$ given by
\be
 \mfu(N)^{+ - }
= \left\{
\begin{bmatrix}
0 & 0 & 0 & 0 \\
0 & 0 & 0 & 0 \\
0 & 0 & 0 & X_{3, 4} \\
0 & 0 & -X_{3, 4}^\dagger & 0
\end{bmatrix}
\, \Bigg| \,
\begin{array}{l}
X_{3, 4} \in \bC^{(s - n) \times n}
\end{array}
\right\},
\ee
\be
 \mfu(N)^{- +}
= \left\{
\begin{bmatrix}
0 & 0 & X_{1, 3} & 0 \\
0 & 0 & X_{2, 3} & 0 \\
-X_{1, 3}^\dagger & -X_{2, 3}^\dagger & 0 & 0 \\
0 & 0 & 0 & 0
\end{bmatrix}
\, \Bigg| \,
\begin{array}{l}
X_{1, 3} \in \bC^{n \times (s - n)},
X_{2, 3} \in \bC^{(r - n) \times (s - n)}
\end{array}
\right\}.
\ee
Note that both $\mfu(N)^{+ -}$ and $\mfu(N)^{- +}$ are trivial if $s = n$.
In general,
$\dim(\mfu(N)^{+ -}) = 2 n (s - n)$ and
$\dim(\mfu(N)^{- +}) = 2 r (s - n)$.
For all $1 \leq j \leq n$ and $1 \leq d \leq s - n$ we define
\be
 \tilde{E}_{\epsilon_j}^{\rr, d}
:= \frac{1}{\sqrt{2}}
\begin{bmatrix}
0 & 0 & 0 & 0 \\
0 & 0 & 0 & 0 \\
0 & 0 & 0 & E_{d j} \\
0 & 0 & - E_{j d} & 0
\end{bmatrix},
\quad
 \tilde{E}_{\epsilon_j}^{\ri, d}
:= \frac{\ri}{\sqrt{2}}
\begin{bmatrix}
0 & 0 & 0 & 0 \\
0 & 0 & 0 & 0 \\
0 & 0 & 0 & E_{d j} \\
0 & 0 & E_{j d} & 0
\end{bmatrix},
\label{3.56}\ee
\be
 \tilde{F}_{\epsilon_j}^{\rr, d}
:= \frac{1}{\sqrt{2}}
\begin{bmatrix}
0 & 0 & - E_{j d} & 0 \\
0 & 0 & 0 & 0 \\
E_{d j} & 0 & 0 & 0 \\
0 & 0 & 0 & 0
\end{bmatrix}, \quad
\tilde{F}_{\epsilon_j}^{\ri, d}
:= \frac{\ri}{\sqrt{2}}
\begin{bmatrix}
0 & 0 & E_{j d} & 0 \\
0 & 0 & 0 & 0 \\
E_{d j} & 0 & 0 & 0 \\
0 & 0 & 0 & 0
\end{bmatrix}.
\label{3.57}\ee
For all $1 \leq c \leq r - n$ and $1 \leq d \leq s - n$ we introduce
\begin{align}
& \tilde{F}_{0}^{\rr, c, d}
:= \frac{1}{\sqrt{2}}
\begin{bmatrix}
0 & 0 & 0 & 0 \\
0 & 0 & E_{c d} & 0 \\
0 & -E_{d c} & 0 & 0 \\
0 & 0 & 0 & 0
\end{bmatrix}, &
& \tilde{F}_{0}^{\ri, c, d}
:= \frac{\ri}{\sqrt{2}}
\begin{bmatrix}
0 & 0 & 0 & 0 \\
0 & 0 & E_{c d} & 0 \\
0 & E_{d c} & 0 & 0 \\
0 & 0 & 0 & 0
\end{bmatrix}. &
\label{3.58}\end{align}
The set of matrices
\be
\{ \tilde{E}_{\epsilon_j}^D \}_{j, D}
:= \{ \tilde{E}_{\epsilon_j}^{\rr, d}, \tilde{E}_{\epsilon_j}^{\ri, d} \}_{\begin{subarray}{l}
1 \leq j \leq n \\
1 \leq d \leq s - n
\end{subarray}}
\label{3.59}\ee
forms an ONB in $\mfu(N)^{+ -}$.
The set of matrices
\be
\{ \tilde{F}_{\epsilon_j}^D \}_{j, D}
:= \{ \tilde{F}_{\epsilon_j}^{\rr, d}, \tilde{F}_{\epsilon_j}^{\ri, d} \}_{\begin{subarray}{l}
1 \leq j \leq n \\
1 \leq d \leq s - n
\end{subarray}}
\label{3.60}\ee
together with the set
\be
\{ \tilde{F}_0^D \}_D
:= \{ \tilde{F}_{0}^{\rr, c, d}, \tilde{F}_{0}^{\ri, c, d} \}_{\begin{subarray}{l}
1 \leq c \leq r - n \\
1 \leq d \leq s - n
\end{subarray}}
\label{3.61}\ee
form an ONB in $\mfu(N)^{- +}$.
They verify the relations
\be
\ad_\bfq(\tilde{E}_{\epsilon_j}^D) = q_j \tilde{F}_{\epsilon_j}^D,
\quad
\ad_\bfq(\tilde{F}_{\epsilon_j}^D) = - q_j \tilde{E}_{\epsilon_j}^D,
\quad
\ad_\bfq(\tilde{F}_0^D) = 0.
\label{commut2}\ee

Now we compute the matrix of $J$ and of $J^{-1}$  on the
 invariant subspaces in (\ref{3.16}).
First, choose an arbitrary ONB $\{ L_j \}_{j = 1}^{ \dim(\cM) }$ in $\cM$.
Then the vectors
\be
\hat{L}_j := \frac{1}{\sqrt{2}} \left( L_j, -L_j \right)
\equiv \frac{1}{\sqrt{2}}
\begin{bmatrix}
L_j \\
-L_j
\end{bmatrix}
\label{hatL}\ee
yield an ONB in $\cK^\perp_a$.
The matrix entries of $J(e^\bfq)|_{\cK^\perp_a}$ and $J(e^\bfq)^{-1}|_{\cK^\perp_a}$
read
\be
\cB_\cG (\hat{L}_k, J(e^\bfq) \hat{L}_l )
= 2 \delta_{k, l},
\quad
\cB_\cG (\hat{L}_k, J(e^\bfq)^{-1} \hat{L}_l )
= \frac{1}{2} \delta_{k, l}.
\label{3.64}\ee
Second, upon introducing the vectors
\be
V_\alpha^D := \frac{1}{\sqrt{2}}
\begin{bmatrix}
E_\alpha^D \\
E_\alpha^D
\end{bmatrix},
\quad
W_\alpha^D := \frac{1}{\sqrt{2}}
\begin{bmatrix}
E_\alpha^D \\
-E_\alpha^D
\end{bmatrix},
\label{3.65}\ee
we obtain an ONB in $\cK^\perp_e$, and by applying
(\ref{J_on_S}) on these vectors we get
\be
J(e^\bfq) V_\alpha^D
= \frac{1}{\sqrt{2}} \begin{bmatrix}
(\1 - \cosh(\ad_\bfq)) E_\alpha^D \\
(\1 - \cosh(\ad_\bfq)) E_\alpha^D
\end{bmatrix},
\quad
J(e^\bfq) W_\alpha^D
= \frac{1}{\sqrt{2}} \begin{bmatrix}
(\1 + \cosh(\ad_\bfq)) E_\alpha^D \\
-(\1 + \cosh(\ad_\bfq)) E_\alpha^D
\end{bmatrix}.
\label{3.66}\ee
We find from the  relations (\ref{commut1}) that
$\cosh(\ad_\bfq) E_\alpha^D = \cos(\alpha(\bfq)) E_\alpha^D$,
and  then elementary trigonometric identities yield
\be
J(e^\bfq) V_\alpha^D
= 2 \sin^2 \left( \frac{\alpha(\bfq)}{2} \right) V_\alpha^D,
\quad
J(e^\bfq) W_\alpha^D
= 2 \cos^2 \left( \frac{\alpha(\bfq)}{2} \right) W_\alpha^D.
\label{3.67}\ee
Therefore the only nontrivial matrix entries of
$J(e^\bfq)|_{\cK^\perp_e}$ and $J(e^\bfq)^{-1}|_{\cK^\perp_e}$
are the following ones:
\begin{align}
& \cB_\cG(V_\alpha^D, J(e^\bfq) V_\alpha^D)
= 2 \sin^2 \left( \frac{\alpha(\bfq)}{2} \right),&
& \cB_\cG(W_\alpha^D, J(e^\bfq) W_\alpha^D)
= 2 \cos^2 \left( \frac{\alpha(\bfq)}{2} \right),& \nonumber \\
& \cB_\cG(V_\alpha^D, J(e^\bfq)^{-1} V_\alpha^D)
= \frac{1}{2 \sin^2 \left( \frac{\alpha(\bfq)}{2} \right)},&
& \cB_\cG(W_\alpha^D, J(e^\bfq)^{-1} W_\alpha^D)
= \frac{1}{2 \cos^2 \left( \frac{\alpha(\bfq)}{2} \right)}.&
\label{3.68}\end{align}
Third, by introducing
\be
\tilde{V}_{\epsilon_j}^D := \frac{1}{\sqrt{2}}
\begin{bmatrix}
\tilde{E}_{\epsilon_j}^D \\
\tilde{F}_{\epsilon_j}^D
\end{bmatrix},
\quad
\tilde{W}_{\epsilon_j}^D := \frac{1}{\sqrt{2}}
\begin{bmatrix}
\tilde{E}_{\epsilon_j}^D \\
-\tilde{F}_{\epsilon_j}^D
\end{bmatrix},
\quad \tilde{Z}_0^D :=
\begin{bmatrix}
0 \\
\tilde{F}_{0}^D
\end{bmatrix},
\label{3.69}\ee
we obtain an ONB in $\cK^\perp_o$, and
the application of (\ref{J_on_S})
on these basis vectors gives
\be
J(e^\bfq) \tilde{V}_{\epsilon_j}^D
= \frac{1}{\sqrt{2}} \begin{bmatrix}
\tilde{E}_{\epsilon_j}^D - \sinh(\ad_\bfq) \tilde{F}_{\epsilon_j}^D \\
\sinh(\ad_\bfq) \tilde{E}_{\epsilon_j}^D + \tilde{F}_{\epsilon_j}^D
\end{bmatrix},
\quad
J(e^\bfq) \tilde{W}_{\epsilon_j}^D
= \frac{1}{\sqrt{2}} \begin{bmatrix}
\tilde{E}_{\epsilon_j}^D + \sinh(\ad_\bfq) \tilde{F}_{\epsilon_j}^D \\
\sinh(\ad_\bfq) \tilde{E}_{\epsilon_j}^D - \tilde{F}_{\epsilon_j}^D
\end{bmatrix}.
\label{3.70}\ee
By using the relations (\ref{commut2}) we see that
\be
J(e^\bfq) \tilde{V}_{\epsilon_j}^D
= (1 + \sin(q_j)) \tilde{V}_{\epsilon_j}^D,
\quad
J(e^\bfq) \tilde{W}_{\epsilon_j}^D
= (1 - \sin(q_j)) \tilde{W}_{\epsilon_j}^D.
\label{3.71}\ee
Since $J(e^\bfq) \tilde{Z}_0^D = \tilde{Z}_0^D$, we conclude that
the only nontrivial matrix entries of $J(e^\bfq)|_{\cK^\perp_o}$ and
its inverse $J(e^\bfq)^{-1}|_{\cK^\perp_o}$ are the following ones:
\begin{align}
& \cB_\cG(\tilde{V}_{\epsilon_j}^D,
J(e^\bfq) \tilde{V}_{\epsilon_j}^D)
= 1 + \sin(q_j),&
& \cB_\cG(\tilde{W}_{\epsilon_j}^D,
J(e^\bfq) \tilde{W}_{\epsilon_j}^D)
= 1 - \sin(q_j),& \nonumber \\
& \cB_\cG(\tilde{V}_{\epsilon_j}^D,
J(e^\bfq)^{-1} \tilde{V}_{\epsilon_j}^D)
= \frac{1}{1 + \sin(q_j)},&
& \cB_\cG(\tilde{W}_{\epsilon_j}^D,
J(e^\bfq)^{-1} \tilde{W}_{\epsilon_j}^D)
= \frac{1}{1 - \sin(q_j)},& \nonumber \\
& \cB_\cG(\tilde{Z}_0^D, J(e^\bfq) \tilde{Z}_0^D) = 1,&
& \cB_\cG(\tilde{Z}_0^D, J(e^\bfq)^{-1} \tilde{Z}_0^D) = 1. &
\label{3.72}\end{align}

\begin{lemma}
By using the identification
$\check \Sigma:=\check A = \exp(\check \cA_+)$ with $\check \cA_+$ in (\ref{cA_check_+}),
the second term of the reduced Laplacian
(\ref{2.14}) is given by
\be
\delta^{-\frac{1}{2}} \Delta_{\check{A}}(\delta^{\frac{1}{2}})
 = \frac{(m - n) (r - s)}{2} \sum_{j = 1}^n \frac{1}{\sin^2(q_j)}
+ \frac{ 4(s - n)^2 - 1}{2} \sum_{j = 1}^n \frac{1}{\sin^2(2 q_j)}
- \frac{n (3 m^2 + n^2 - 1)}{6}.
\label{mes}\ee
\end{lemma}
\textbf{Proof.}
Consider the function
\be
\cJ :=
\prod_{1 \leq k < l \leq n} \left[ \sin(q_k - q_l) \sin(q_k + q_l) \right]^\nu
\prod_{j = 1}^n \left[ \sin(q_j) \right]^{\nu_1}
\prod_{j = 1}^n \left[ \sin(2 q_j) \right]^{\nu_2},
\label{B1}\ee
where the domain of the variables $q_1, q_2, \ldots, q_n$ is such that all
$\sin$ functions are positive  and
$\nu, \nu_1, \nu_2 \in \bR$ are arbitrary parameters.
Recall from \cite{OPfa} the identity
\be\begin{split}
\cJ^{-1} \sum_{a = 1}^n \frac{\partial^2  \cJ}{\partial q_a^2} =
& \nu (\nu - 1) \sum_{1 \leq k < l \leq n}
\left( \frac{1}{\sin^2(q_k - q_l)} + \frac{1}{\sin^2(q_k + q_l)}  \right) \\
& + \nu_1 (\nu_1 + 2 \nu_2 - 1) \sum_{j = 1}^n \frac{1}{\sin^2(q_j)}
+ 4 \nu_2 (\nu_2 - 1) \sum_{j = 1}^n \frac{1}{\sin^2(2 q_j)} \\
& - n \left[ (\nu_1 + 2 \nu_2)^2  + 2 \nu (\nu_1 + 2 \nu_2)  (n - 1)
+\frac{2}{3} \nu^2 (n - 1) (2 n - 1) \right].
\end{split}
\label{B2}\ee
By calculating $\det (J(e^{\bfq}))$ using the above
basis of $\cK^\perp$, it is easily obtained from (\ref{2.13}) that
$\delta^{\frac{1}{2}}(e^\bfq) \propto \cJ(q_1,q_2,\ldots, q_n)$ with
\be
\nu=1, \quad \nu_1=r - s,\quad \nu_2= s - n + \frac{1}{2}.
\ee
Taking into account  (\ref{LB_on_checkA}),
the required statement follows immediately.
\hspace*{\stretch{1}} \qedsymb
\medskip

The subsequent formula is obtained by direct substitution since we
have determined  the matrix elements of $J(e^\bfq)^{-1}$ (cf.~(\ref{b-def})).
It will be used in Section 4, when we
shall further inspect the reduced Laplace operator (\ref{2.14}) in interesting cases.

\begin{lemma}
In terms of the above notations,
the third term of the reduced Laplacian (\ref{2.14}) takes the following form:
\begin{align}
b^{\alpha, \beta} \rho\prime(T_\alpha) \rho\prime(T_\beta)
= & \frac{1}{2} \sum_{1 \leq j \leq \dim(\cM)} \rho\prime(\hat{L}_j)^2
+ \frac{1}{2} \sum_{j = 1}^n \left(
\frac{\rho\prime(V_{2 \epsilon_j}^\ri)^2}{\sin^2 (q_j)}
+ \frac{\rho\prime(W_{2 \epsilon_j}^\ri)^2}{\cos^2 (q_j)} \right) \nonumber \\
& + \frac{1}{2} \sum_{1 \leq k < l \leq n} \left(
\frac{ \rho \prime (V_{\epsilon_k - \epsilon_l}^{\rr})^2 +
\rho\prime(V_{\epsilon_k - \epsilon_l}^{\ri})^2}
{ \sin^2 \left( \frac{q_k - q_l}{2} \right) }
+ \frac{\rho\prime(W_{\epsilon_k - \epsilon_l}^\rr)^2 +
\rho\prime(W_{\epsilon_k - \epsilon_l}^\ri)^2}
{\cos^2 \left(\frac{q_k - q_l}{2}\right)} \right) \nonumber \\
& + \frac{1}{2} \sum_{1 \leq k < l \leq n} \left(
\frac{ \rho \prime (V_{\epsilon_k + \epsilon_l}^{\rr})^2 + \rho\prime(V_{\epsilon_k + \epsilon_l}^{\ri})^2}
{ \sin^2 \left( \frac{q_k + q_l}{2} \right) }
+ \frac{\rho\prime(W_{\epsilon_k + \epsilon_l}^\rr)^2 + \rho\prime(W_{\epsilon_k + \epsilon_l}^\ri)^2}
{\cos^2 \left(\frac{q_k + q_l}{2}\right)} \right) \nonumber \\
& + \frac{1}{2} \sum_{j = 1}^n \sum_{d = 1}^{r - n} \left(
\frac{ \rho \prime (V_{\epsilon_j}^{\rr, d})^2 + \rho\prime(V_{\epsilon_j}^{\ri, d})^2}
{ \sin^2 \left( \frac{q_j}{2} \right) }
+ \frac{\rho\prime(W_{\epsilon_j}^{\rr, d})^2 + \rho\prime(W_{\epsilon_j}^{\ri, d})^2}
{\cos^2 \left(\frac{q_j}{2}\right)} \right) \nonumber \\
& + \sum_{j = 1}^n \sum_{d = 1}^{s - n} \left(
\frac{ \rho \prime (\tilde{V}_{\epsilon_j}^{\rr, d})^2 + \rho\prime(\tilde{V}_{\epsilon_j}^{\ri, d})^2}
{ 1 + \sin(q_j) }
+ \frac{\rho\prime(\tilde{W}_{\epsilon_j}^{\rr, d})^2 + \rho\prime(\tilde{W}_{\epsilon_j}^{\ri, d})^2}
{1 - \sin(q_j)} \right) \nonumber \\
& + \sum_{c = 1}^{r - n} \sum_{d = 1}^{s - n}
\left( \rho\prime(\tilde{Z}_0^{\rr, c, d})^2 + \rho\prime(\tilde{Z}_0^{\ri, c, d})^2 \right).
\label{4.41}\end{align}
\end{lemma}

\section{$BC_n$ Sutherland models from the KKS ansatz}
\setcounter{equation}{0}

In this section we study interesting examples of the quantum Hamiltonian reduction
based on
the Hermann action (3.2) on $Y=U(N)$ associated with the involutions (\ref{3.30}).
The reductions correspond to certain UIRREPS $\rho$ of the symmetry group
\be
G = U(N)_L \times U(N)_R = (U(r) \times U(s)) \times (U(m) \times U(n)).
\label{4.1}\ee
To describe them,  we now briefly summarize our notations for the UIRREPS of $U(n)$, for
arbitrary $n$.
(See also Appendix A.)
First, we have the UIRREP  $(\bfPi_\lambda, V_\lambda)$ of $SU(n)$ in correspondence to
any  highest weight
$\lambda \in P_+(SU(n))$, that can be written as
$\lambda=\sum_{i=1}^{n-1} a_i \varpi_i$ using the fundamental weights
$\varpi_i$ and integers $a_i\in \bZ_{\geq 0}$.
A label
$\mu_n(\lambda) \in \{0,1,\ldots, n-1\}$ is  attached to the highest weight
$\lambda$ by the congruence relation
\be
\mu_n(\lambda) \equiv  \sum_{k=1}^{n-1} k a_k  \pmod n
\quad\hbox{for}\quad
\lambda=\sum_{i=1}^{n-1} a_i \varpi_i.
\label{mu_n_value}\ee
It enters the equality
$\bfPi_\lambda( e^{\ri \frac{2 \pi}{n}} \1_n)
= e^{\ri \frac{2 \pi}{n} \mu_n(\lambda)} \Id_{V_\lambda}$.
Then, for any $k\in \bZ$, the representation $\bfPi_\lambda$ of $SU(n)$ extends
 to the representation $\rho_{(k,\lambda)}$ of $U(n)$ defined by
\be
\rho_{(k,\lambda)}(\xi g) = \xi^{nk + \mu_n(\lambda)} \bfPi_\lambda(g),
\qquad
\forall \xi\in U(1),\,\forall g\in SU(n).
\label{Urep}\ee
Up to equivalence, all UIRREPS of $U(n)$ are obtained in this way.
The notation makes sense even
for $n=1$, by putting $P_+(SU(1)):=\{0\}$, and we have
 $\rho_{(k,0)}(g) = (\det g)^k$ ($\forall g\in U(n)$).
 By letting
 $\rho\prime_{(k,\lambda)}$ and $\pi_\lambda$ stand for the infinitesimal version
 of the representations $\rho_{(k, \lambda)}$ and $\bfPi_\lambda$, respectively, we have
\be
\rho\prime_{(k, \lambda)}(Z)=
\pi_\lambda \left( Z - \frac{\tr(Z)}{n} \1_n \right)
+ (\mu_n(\lambda) + n k) \frac{\tr(Z)}{n} \Id_{V_\lambda},
\qquad \forall Z \in \mfu(n).
\label{rhoprime+}
\ee
We use the notations
$\pi_\lambda^{(n)}$, $V^{(n)}_\lambda$, $\rho_{(k,\lambda)}^{(n)}$ etc.~when considering
various values of $n$ simultaneously.

The UIRREPS of the direct product group $G$ (\ref{4.1})
have the form
\be
\rho = \left(
\rho^{(r)}_{(k_L^1, \lambda_L^1)} \boxtimes \rho^{(s)}_{(k_L^2, \lambda_L^2)}
\right)
\boxtimes
\left(
\rho^{(m)}_{(k_R^1, \lambda_R^1)} \boxtimes \rho^{(n)}_{(k_R^2, \lambda_R^2)}
\right),
\label{4.2}\ee
where $\lambda_L^1, \lambda_L^2, \lambda_R^1, \lambda_R^2$ are the highest weights
and $k_L^1, k_L^2, k_R^1, k_R^2 \in \bZ$ according to (\ref{Urep}).
The main problem is to find the UIRREPS $(\rho, V)$ for which
\be
\dim (V^K) = 1,
\label{4.3}\ee
where $K= M_\diag < G$ is given by (\ref{3.41}).
 We  investigate this problem
by adopting the ansatz that one of the 4 constituent representations in (\ref{4.2}) has
the form $\rho^{(l)}_{(k, a_1\varpi_1)}$ ($l\in \{r,s,m,n\}$)
and the other 3 constituent representations are one-dimensional.
More exactly, $\rho^{(l)}_{(k, a_1\varpi_1)}$ will be used for a factor of the maximal size,
$l=\max\{r,s,m,n\}$.
We call this assumption the
\emph{KKS ansatz}, since it eventually originates from the seminal paper by Kazhdan,
Kostant and Sternberg \cite{KKS}.
The usefulness of this assumption is also supported by results in \cite{EFK,Obl,LMP}.
The key property is that all weight-multiplicities
of $\rho^{(l)}_{(k, a_1\varpi_1)}$ are equal to one.
The analysis of the condition (\ref{4.3}) is the easiest
if the group $K$ (\ref{3.41}) is Abelian, which happens in the following cases:
\begin{itemize}
\item case I: $m = r = s = n$, $N = 2 n$,
\item case II: $m = r = n + 1$, $s = n$, $N = 2 n + 1$,
\item case III: $m = n + 2$, $r = s = n + 1$, $N = 2 n + 2$.
\end{itemize}
Next we describe the simplest case I
in detail,  then  present the essential points for the other two cases.
The complex holomorphic analogue of case I was studied in \cite{Obl};
and the results are consistent.
The other two cases of our KKS ansatz have not been investigated before.

\medskip
\noindent
\textbf{Remark:}
The reader may wonder why we take $l=\max\{r,s,m,n\}$ in our KKS ansatz in cases II and III.
In fact, we previously  studied (\cite{LMP} and unpublished work) the classical Hamiltonian  reductions
of the free particle on $U(N)$ based on the symmetry group (\ref{4.1}) by using
a minimal coadjoint orbit of positive  dimension for \emph{any} one of the $4$ factors
and one-point orbits for the other 3 factors.
We found that this leads to the classical $BC_n$ Sutherland model with three independent
coupling constants \emph{only} in the three  cases mentioned above, and \emph{only} if
the minimal coadjoint orbit of positive dimension, $2(l-1)$ for $U(l)$,  is associated with a factor of
maximal size.
The connection to quantum Hamiltonian reduction is clear from the relation
between the
coadjoint orbits of $U(l)$ of dimension $2(l-1)$ and the representations
$\rho^{(l)}_{(k,a_1\varpi_1)}$ (and their contragredients), which follows for example
from geometric quantization.

\subsection{Case I: $m = r = s = n$, $N = 2 n$}

Now
$\sigma_L = \sigma_R = \theta_{n, n}$ and $U(N)_L = U(N)_R \cong U(n) \times U(n)$.
The decomposition (\ref{matrixDecomp}) of any matrix in
$\bC^{N \times N}$  simplifies to a two by two block form  with all 4 blocks
having size $n \times n$.
We look for admissible UIRREPS $\rho$ of
$G$ (\ref{4.1}) by adopting the KKS ansatz
\be
\rho := \left(
\rho^{(n)}_{(k_L^1, a_1 \varpi_1)} \boxtimes \rho^{(n)}_{(k_L^2, 0)}
\right)
\boxtimes
\left(
\rho^{(n)}_{(k_R^1, 0)} \boxtimes \rho^{(n)}_{(k_R^2, 0)}
\right),
\label{4.5}\ee
where $a_1 \in \bZ_{\geq 0}$, $k_L^1, k_L^2, k_R^1, k_R^2 \in \bZ$ and
the representation space  is identified as
\be
V \equiv  V_{a_1 \varpi_1}^{(n)}.
\label{identification_I}\ee
Note that any element
$X \in \cG \cong \mfu(N)^{\sigma_L, +} \oplus \mfu(N)^{\sigma_R, +}$
of the symmetry algebra $\cG$ can be realized as a pair $X = (X_L, X_R)$ with
$X_L, X_R \in \mfu(N)^{\sigma_L, +} = \mfu(N)^{\sigma_R, +}
\cong \mfu(n) \oplus \mfu(n)$.
So, for any $X \in \cG$ we have the  refined decomposition
\be
X = (X_L, X_R)
= \left( (X_L^1, X_L^2), (X_R^1, X_R^2) \right),
\label{4.7}\ee
where  $X_L^1, X_L^2, X_R^1, X_R^2 \in \mfu(n)$ and as block-matrices
\be
(X_L^1, X_L^2):=
\begin{bmatrix}
X_L^1 & 0 \\
0 & X_L^2
\end{bmatrix},
\qquad
(X_R^1, X_R^2):=
\begin{bmatrix}
X_R^1 & 0 \\
0 & X_R^2
\end{bmatrix}.
\label{4.8}\ee
With these notations, the formula of the Lie algebra representation
corresponding to (\ref{4.5}) reads
\be
\rho\prime(X)
=\pi_{a_1 \varpi_1}^{(n)} \bigl(
X_L^1 - \frac{\tr(X_L^1)}{n} \1_n\bigr)
+ \left[\bigl(k_L^1 + \frac{\mu_n(a_1 \varpi_1)}{n}\bigr)\tr(X_L^1) + \tr(k_L^2 X_L^2
+ k_R^1 X_R^1 + k_R^2 X_R^2)\right] \Id_{V}.
\label{rhoprime_I}
\ee
\begin{lemma}
The KKS ansatz (\ref{4.5}) defines admissible UIRREPS of $G$
 satisfying $\dim(V^K)\neq 0$  if and only if
$k_L^1 + k_L^2 + k_R^1 + k_R^2 = 0$ and $a_1 = \gamma n$
with some $\gamma \in \bZ_{\geq 0}$.  In these cases
$\dim(V^K) = 1$.
Using the bosonic oscillator realization of $V$ (\ref{identification_I}) described in Appendix A,
 $V^K$ has
the form
\be
V^K \cong V_{\gamma n \varpi_1}^{(n)}[0] =
\Span_\bC \{ \ket{ \gamma, \gamma, \ldots, \gamma } \}.
\label{4.10}\ee
\end{lemma}
\textbf{Proof.}
The isotropy subalgebra is
$\cK = \cM_\diag = \{ X = (X_0, X_0) \, | \, X_0 \in \cM \}$,
where $\cM$ can be parametrized as
\be
\cM = \left\{
X_0 = \begin{bmatrix}
H + \ri x \1_n & 0 \\
0 & H + \ri x \1_n
\end{bmatrix}
\, \bigg| \,
H \in \ri \cH_\bR^{(n)}, x \in \bR \right\}.
\ee
That is, for the components of any $X \in \cK$ we have the
parametrization
\be
X_L^1 = X_L^2 = X_R^1 = X_R^2 = H + \ri x \1_n.
\ee
Thus, using equation (\ref{rhoprime_I}), for any
$v \in V_{a_1 \varpi_1}^{(n)}$ and $X\in \cK$ we can write
\be
\rho\prime(X) v
= \pi_{a_1 \varpi_1}^{(n)}(H) v
+ \ri x \left( \mu_n(a_1 \varpi_1)
+ n (k_L^1 + k_L^2 + k_R^1 + k_R^2)\right) v.
\ee
Clearly $\rho\prime(X) v = 0$ $(\forall X \in \cK)$
if and only if
\be
\pi_{a_1 \varpi_1}^{(n)}(H) v = 0
\quad
(\forall H \in \ri \cH_\bR^{(n)})
\quad \mbox{and} \quad
\mu_n(a_1 \varpi_1) + n (k_L^1 + k_L^2 + k_R^1 + k_R^2) = 0.
\label{4.14}\ee
Therefore $V^K = V^\cK \cong V_{a_1 \varpi_1}^{(n)}[0]$,
provided that
$\mu_n(a_1 \varpi_1) + n (k_L^1 + k_L^2 + k_R^1 + k_R^2) = 0$.
It is easy to see that
$V_{a_1 \varpi_1}^{(n)}[0]\neq \{0\}$
if and only if $a_1 = \gamma n$ for some $\gamma \in \bZ_{\geq 0}$.
Since $\mu_n(\gamma n \varpi_1) = 0$ by  (\ref{mu_n_value}),
the  requirement $k_L^1 + k_L^2 + k_R^1 + k_R^2 = 0$
then also follows from (\ref{4.14}).
Finally, note that by using  the oscillator realization of
 $V_{\gamma n \varpi_1}^{(n)}$ one  has the second equality in (\ref{4.10}).
\hspace*{\stretch{1}} \qedsymb
\medskip

In what follows we  make use of the basis of $\cK^\perp$ constructed
in subsection 3.3. In the present case this is given by the basis
$\{ V_\alpha^a, W_\alpha^a \}_{a \in \{ \rr, \ri \},
\alpha \in \cR_+(C_n)}$ of
$\cK^\perp_e$ together with the basis
$\{\hat{L}_j\}$ of $\cK^\perp_a$ defined according to (\ref{hatL}) by using
the following orthonormal
basis $\{ L_j \}_{j = 1}^n$ of $\cM$:
\be
L_j := \frac{\ri}{\sqrt{2}}
\begin{bmatrix}
E_{jj} & 0 \\
0 & E_{jj}
\end{bmatrix}
\in \cM
\quad
(1 \leq j \leq n).
\ee

\begin{lemma}
In the case of the KKS ansatz (\ref{4.5}) subject to the conditions of Lemma 4.1
the third term in the reduced Laplacian (2.14) gives
\bea
&& b^{\alpha, \beta} \rho\prime(T_\alpha) \rho\prime(T_\beta)
=  - \frac{1}{2} n (k_L^1 + k_L^2)^2
- \gamma ( \gamma + 1 ) \sum_{1 \leq k < l \leq n} \left(
\frac{1}{\sin^2(q_k - q_l)} + \frac{1}{\sin^2(q_k + q_l)} \right) \nonumber \\
&& \phantom{XXXXX}  - \frac{(k_L^1 + k_R^1)^2 - (k_L^2 + k_R^1)^2}{2}
\sum_{j = 1}^n \frac{1}{\sin^2(q_j)}
- 2 (k_L^2 + k_R^1)^2 \sum_{j = 1}^n \frac{1}{\sin^2(2 q_j)}.
\label{4.42}\eea
\end{lemma}
\textbf{Proof.}
 Note that in the present case
only the first 4 sums occur in the formula (\ref{4.41}).
Recalling that $\mu_n(\gamma n \varpi_1) = 0$ and utilizing formula
(\ref{rhoprime_I}) for $\rho\prime$, we can calculate
the action of the various terms.
 For example, since
\be
\hat{L}_j = \frac{1}{\sqrt{2}}(L_j, - L_j)
= \frac{\ri}{2} \left( (E_{jj}, E_{jj}), (- E_{jj}, - E_{jj}) \right),
\ee
we get
\be
\rho\prime(\hat{L}_j)
= \frac{\ri}{2}
\left( \pi_{\gamma n \varpi_1}^{(n)}\left(E_{jj} - \frac{1}{n}\1_n\right)
+ (k_L^1 + k_L^2 - k_R^1 - k_R^2) \Id_V \right).
\ee
The action of $\rho\prime(\hat{L}_j)$ on $V^K$ can be easily calculated
in the bosonic oscillator picture. Since
$\pi_{\gamma n \varpi_1}^{(n)}\left(E_{jj} - \frac{1}{n}\1_n\right)
\ket{\gamma, \gamma, \ldots, \gamma}
= 0$,
and since $k_R^2 = -k_L^1 - k_L^2 - k_R^1$, it follows that on
the subspace $V^K \cong \Span_\bC\{\ket{\gamma, \gamma, \ldots, \gamma}\}$
the operator $\rho\prime(\hat{L}_j)$ acts as the scalar
$\rho\prime(\hat{L}_j) = \ri( k_L^1 + k_L^2 )$.
In the same manner,  the equalities
$\rho\prime(V_{2 \epsilon_j}^\ri) = \ri (k_L^1 + k_R^1)$ and
$\rho\prime(W_{2 \epsilon_j}^\ri) = - \ri (k_L^2 + k_R^1)$ hold on $V^K$.
Furthermore, we have on $V$
\begin{align}
& \rho\prime(V_{\epsilon_k - \epsilon_l}^\rr)
= \rho\prime(W_{\epsilon_k - \epsilon_l}^\rr)
= \rho\prime(V_{\epsilon_k + \epsilon_l}^\rr)
= \rho\prime(W_{\epsilon_k + \epsilon_l}^\rr)
= \frac{1}{2 \sqrt{2}}\left( \pi_{\gamma n \varpi_1}^{(n)}(E_{kl})
- \pi_{\gamma n \varpi_1}^{(n)}(E_{lk}) \right), \\
& \rho\prime(V_{\epsilon_k - \epsilon_l}^\ri)
= \rho\prime(W_{\epsilon_k - \epsilon_l}^\ri)
= \rho\prime(V_{\epsilon_k + \epsilon_l}^\ri)
= \rho\prime(W_{\epsilon_k + \epsilon_l}^\ri)
= \frac{\ri}{2 \sqrt{2}}\left( \pi_{\gamma n \varpi_1}^{(n)}(E_{kl})
+ \pi_{\gamma n \varpi_1}^{(n)}(E_{lk}) \right).
\end{align}
Next,  $\forall k, l \in \{1, 2, \ldots, n\}$, $k \neq l$,
we obtain
\be
\pi_{\gamma n \varpi_1}^{(n)}(E_{k l}) \pi_{\gamma n \varpi_1}^{(n)}(E_{l k})
\ket{\gamma, \gamma, \ldots, \gamma}
 = b_k^\dagger b_l b_l^\dagger b_k \ket{\gamma, \gamma, \ldots, \gamma}
= \gamma ( \gamma + 1 ) \ket{\gamma, \gamma, \ldots, \gamma}.
\ee
The above equations imply that on $V^K$
\be
\rho\prime(\hat{L}_j)^2 = -(k_L^1 + k_L^2)^2,
\quad
\rho\prime(V_{2 \epsilon_j}^\ri)^2 = -(k_L^1 + k_R^1)^2,
\quad
\rho\prime(W_{2 \epsilon_j}^\ri)^2 = -(k_L^2 + k_R^1)^2,
\ee
\be
\rho\prime (V_{\alpha}^\rr)^2
+ \rho\prime(V_{\alpha}^\ri)^2
= \rho\prime(W_{\alpha}^\rr)^2
+ \rho\prime(W_{\alpha}^\ri)^2
=  -\frac{1}{2} \gamma ( \gamma + 1 )
\quad
\hbox{for}\quad  \alpha = \epsilon_k \pm \epsilon_l,\,\, k\neq l.
\ee
Now (\ref{4.42}) results by substitution into (\ref{4.41}), using
obvious trigonometric identities.
\hspace*{\stretch{1}} \qedsymb

\medskip
The following proposition is obtained by putting together the statements of
equation (\ref{LB_on_checkA}),  Lemma 3.2 and Lemma 4.2.

\begin{proposition}
Under the KKS ansatz (\ref{4.5}) the general formula (\ref{2.14}) gives
the following result for the reduction of the Laplace operator of $U(N)$:
\be
-\Delta_\red =H_{BC_n} +
\frac{1}{2} n (k_L^1 + k_L^2)^2 - \frac{1}{6} n (2 n - 1)(2 n + 1),
\ee
where $H_{BC_n}$ is the Sutherland Hamiltonian (\ref{**}) with the coupling parameters
defined by
\be
a \equiv \gamma,
\quad
b \equiv \vert k_L^1 + k_R^1\vert,
\quad
c \equiv \vert k_L^2 + k_R^1\vert
\ee
in terms of the free parameters
$k_L^1, k_L^2, k_R^1 \in \bZ$ and $\gamma \in \bZ_{\geq 0}$
determined by Lemma 4.1.
\end{proposition}
\noindent {\bf Remark:}
By varying $\gamma, k_L^1, k_L^2, k_R^1$,
the coupling parameters $a, b, c$ in (\ref{**}) can take arbitrary non-negative integer values.
As further discussed in Section 5,
 Proposition 4.3 follows also from the results of Oblomkov \cite{Obl}.

\subsection{Case II: $m = r = n + 1$, $s = n$, $N = 2 n + 1$}

In this case $\sigma_L = \sigma_R = \theta_{n + 1, n}$ and correspondingly
$U(N)_L = U(N)_R \cong U(n + 1) \times U(n)$.
We consider the following ansatz for the UIRREP $(\rho,V)$ of the symmetry
group $G$ (\ref{4.1}),
\be
\rho:= \left(\rho^{(n + 1)}_{(k_L^1, a_1\varpi_1)} \boxtimes \rho^{(n)}_{(k_L^2, 0)}\right)
\boxtimes
\left(
\rho^{(n + 1)}_{(k_R^1, 0)} \boxtimes \rho^{(n)}_{(k_R^2, 0)}\right),
\label{KKSII}\ee
where $a_1\in \bZ_{\geq 0}$,  $k_L^1, k_L^2, k_R^1, k_R^2 \in \bZ$
and the carrier space is identified as
$V\equiv V_{a_1 \varpi_1}^{(n+1)}$.
Similarly to  (\ref{4.7}), any
$X \in \cG \cong \mfu(N)^{\sigma_L, +} \oplus \mfu(N)^{\sigma_R, +}$
can be realized as a pair $X = (X_L, X_R)$ with
$X_L, X_R \in \mfu(N)^{\sigma_L, +} = \mfu(N)^{\sigma_R, +}
\cong \mfu(n + 1) \oplus \mfu(n)$.
So, we write $X \in \cG$ as
$X = (X_L, X_R)
= \left( (X_L^1, X_L^2), (X_R^1, X_R^2) \right)$
with $X_L^1, X_R^1 \in \mfu(n + 1)$, $X_L^2, X_R^2 \in \mfu(n)$.
Then (\ref{KKSII})  implies the formula
\bea
&& \rho\prime(X)=\pi_{a_1 \varpi_1}^{(n + 1)} \left(
X_L^1 - \frac{\tr(X_L^1)}{n + 1} \1_{n + 1} \right) \nonumber\\
&&
\phantom{X}+ \left[\left( k_L^1 + \frac{\mu_{n + 1}(a_1 \varpi_1)}{n + 1} \right)
\tr(X_L^1) +  k_L^2 \tr(X_L^2)
+ k_R^1 \tr(X_R^1) + k_R^2 \tr(X_R^2)\right]\Id_V.
\label{rhoprime_II}
\eea
\begin{lemma}
The KKS ansatz (\ref{KKSII}) yields admissible UIRREPS of $G$ if and only if
$\exists \gamma, \tilde{\gamma} \in \bZ_{\geq 0}$ such that
the parameters $k_L^1, k_L^2, k_R^1, k_R^2 \in \bZ$, and
$a_1 \in \bZ_{\geq 0}$ satisfy the conditions
\be
a_1 = \gamma n + \tilde{\gamma},
\quad
k_L^2 + k_R^2 = \tilde{\gamma} - \gamma,
\quad
k_L^1 + k_R^1 = R - (\tilde{\gamma} - \gamma),
\label{cond_II}
\ee
where $\tilde{\gamma} - \gamma = Q + (n + 1) R$
with uniquely determined
$Q = Q(\gamma, \tilde{\gamma}) \in \{ 0, 1, \ldots, n \}$
and $R = R(\gamma, \tilde{\gamma}) \in \bZ$.
If these conditions hold, then
$\dim(V^K) = 1$ and
$V^K$ is given by
\be
V^K \cong
V_{a_1 \varpi_1}^{(n + 1)}[ \gamma e_1 + \gamma e_2 + \cdots + \gamma e_n + \tilde{\gamma} e_{n + 1} ]
= \Span_\bC \{ \ket{ \gamma, \gamma, \ldots, \gamma, \tilde{\gamma} } \},
\ee
where the last equality refers to the bosonic oscillator realization of $V_{a_1 \varpi_1}^{(n + 1)}$.
\end{lemma}
\textbf{Proof.}
For the isotropy subalgebra we have
$\cK = \cM_\diag = \{ X = (X_0, X_0) \, | \, X_0 \in \cM \}$,
where
\be
\cM = \left\{
X_0 = \ri \begin{bmatrix}
D & 0 & 0 \\
0 & \omega & 0 \\
0 & 0 & D
\end{bmatrix}
\, \bigg| \,
D = \diag(d_1, d_2, \ldots, d_n) \in \bR^{n \times n}, \omega \in \bR
\right\}.
\ee
So, for
any $X \in \cK$ we have $X_L = X_R = X_0$, and
\be
X_L^1 = X_R^1 = \ri \begin{bmatrix}
D & 0 \\
0 & \omega
\end{bmatrix},
\quad
X_L^2 = X_R^2 = \ri D.
\ee
Now, for each $\varphi = (\varphi_1, \varphi_2, \ldots, \varphi_n) \in \bR^n$
we let $\bar{\varphi} := \sum_{j = 1}^n \varphi_j$, and consider
the traceless Cartan elements
\be
H_\varphi := \diag(\varphi_1, \varphi_2, \ldots, \varphi_n, - \bar{\varphi})
\in \cH^{(n+1)}_{\bR},
\quad
\tilde{H}_\varphi := \diag(\varphi_1, \varphi_2, \ldots, \varphi_n)
- \frac{1}{n} \bar{\varphi} \1_n \in \cH_{\bR}^{(n)}.
\ee
Then the components of $X \in \cK$ can be
parametrized as
\be
X_L^1 = X_R^1 = \ri H_\varphi + \ri x \1_{n + 1},
\quad
X_L^2 = X_R^2 = \ri \tilde{H}_\varphi
+ \ri \left( x + \frac{1}{n} \bar{\varphi} \right) \1_n,
\ee
where $\varphi \in \bR^n$ and $x \in \bR$. From (\ref{rhoprime_II})
it follows that $\forall v \in V_{a_1 \varpi_1}^{(n + 1)}$ we have
\be
\rho\prime(X) v
= \pi_{a_1 \varpi_1}^{(n + 1)}(\ri H_\varphi) v
+ \ri (k_L^2 + k_R^2) \bar{\varphi} v
+ \ri x \left(
\mu_{n + 1}(a_1 \varpi_1) + (n + 1)(k_L^1 + k_R^1)
+ n(k_L^2 + k_R^2) \right) v.
\ee
Clearly $\rho\prime(X) v = 0$ $(\forall X \in \cK)$
if and only if
\be
\pi_{a_1 \varpi_1}^{(n + 1)}(H_\varphi) v = - (k_L^2 + k_R^2) \bar{\varphi} v
\quad
(\forall \varphi \in \bR^n),
\ee
and $\mu_{n + 1}(a_1 \varpi_1) + (n + 1)(k_L^1 + k_R^1)
+ n(k_L^2 + k_R^2) = 0$.
Note that
$\bar{\varphi} = \sum_{j = 1}^n \varphi_j = \sum_{j = 1}^n e_j(H_\varphi)$,
so after introducing the shorthand notations
\be
\kappa_1 := k_L^1 + k_R^1 \in \bZ
\quad
\mbox{and}
\quad
\kappa_2 := k_L^2 + k_R^2 \in \bZ,
\ee
we conclude that
\be
V^K = V^\cK \cong V_{a_1 \varpi_1}^{(n + 1)}[ - \kappa_2 \sum_{j = 1}^n e_j ],
\ee
provided that $\mu_{n + 1}(a_1 \varpi_1) + (n + 1) \kappa_1 + n \kappa_2 = 0$.
Our next goal is to identify the weight space
$V_{a_1 \varpi_1}^{(n + 1)}[- \kappa_2 (e_1 + e_2 + \cdots + e_n) ]$.
Recall that $-\kappa_2 (e_1 + e_2 + \cdots + e_n) \in \cW_{a_1 \varpi_1}^{(n + 1)}$
if and only if $\exists (l_1, l_2, \ldots, l_{n + 1}) \in \bZ_{\geq 0}^{n + 1}$
with $l_1 + l_2 + \cdots + l_{n + 1} = a_1$, such that
\be
-\kappa_2 (e_1 + e_2 + \cdots + e_n) = \sum_{j = 1}^{n + 1} l_j e_j
= \sum_{j = 1}^n (l_j - l_{n + 1}) e_j.
\ee
Since the functionals
$e_1, e_2, \ldots, e_n$ are linearly independent, we end up with
the requirement
$l_1 = l_2 = \ldots = l_n = l_{n + 1} - \kappa_2$.
For the free parameters we choose $\gamma := l_1$ and
$\tilde{\gamma} := l_{n + 1}$, then the parameters $\kappa_2 = k_L^2 + k_R^2$
and $a_1$ have to obey the equations $\kappa_2 = \tilde{\gamma} - \gamma$
and $a_1 = \gamma n + \tilde{\gamma}$. Note that under these
assumptions we have
\be
V_{a_1 \varpi_1}^{(n + 1)}[- \kappa_2 (e_1 + e_2 + \cdots + e_n) ]
= V_{a_1 \varpi_1}^{(n + 1)}
[ \gamma e_1 + \gamma e_2 + \cdots + \gamma e_n + \tilde{\gamma} e_{n + 1} ]
= \Span_\bC \{ \ket{\gamma, \gamma, \ldots, \gamma, \tilde{\gamma}} \}.
\ee
Now let us express the value of the label
$\mu_{n + 1}(a_1 \varpi_1) \in \{ 0, 1, \ldots, n \}$
in terms of $\gamma$ and $\tilde{\gamma}$. Recalling
(\ref{mu_n_value}), we can write
\be
\mu_{n + 1}(a_1 \varpi_1)
= \mu_{n + 1}((\gamma n + \tilde{\gamma}) \varpi_1)
\equiv \gamma n + \tilde{\gamma}
\equiv \tilde{\gamma} - \gamma
\pmod{(n+1)}.
\ee
Notice that
$\exists!\, Q = Q(\gamma, \tilde{\gamma}) \in \{ 0, 1, \ldots , n \}$
and
$\exists!\, R = R(\gamma, \tilde{\gamma}) \in \bZ$
such that $\tilde{\gamma} - \gamma = Q + (n + 1) R$,
thereby the previous congruence relation translates into the
equation $\mu_{n + 1}(a_1 \varpi_1) = Q$. Plugging this
equation into the requirement
$\mu_{n + 1}(a_1 \varpi_1) + (n + 1) \kappa_1 + n \kappa_2 = 0$,
we get
\be
0 = Q + (n + 1) \kappa_1 + n \left( Q + (n + 1) R \right)
=(n + 1)(\tilde{\gamma} - \gamma - R + \kappa_1),
\ee
therefore we end up with the additional constraint
$k_L^1 + k_R^1 = \kappa_1 = R - (\tilde{\gamma} - \gamma)$.
\hspace*{\stretch{1}} \qedsymb

\medskip
Observe from Lemma 4.4 that
 $k_R^1, k_R^2 \in \bZ$ and $\gamma, \tilde{\gamma} \in \bZ_{\geq 0}$ can be taken
as \emph{free parameters} that label the admissible cases of the KKS ansatz (\ref{KKSII}).
By proceeding like in subsection 4.1, it is matter
of straightforward substitutions to specialize the reduced Laplacian (\ref{2.14}) to our case.
 In this way  we found the following result.

\begin{proposition}
Under the KKS ansatz (\ref{KKSII}) with parameters satisfying
(\ref{cond_II}) the Laplace operator of $U(N)$ reduces to
\be
-\Delta_\red =H_{BC_n} +
\frac{1}{2} n (k_R^1 + k_R^2)^2 + (k_R^1)^2 - \frac{1}{3} n (n + 1)(2 n + 1),
\ee
where $H_{BC_n}$ is given by (\ref{**}) with the coupling parameters determined
in terms of the  arbitrary parameters
 $k_R^1, k_R^2 \in \bZ$ and $\gamma, \tilde{\gamma} \in \bZ_{\geq 0}$
 according to
\be
a \equiv \gamma,
\quad
b \equiv\gamma + \tilde{\gamma} + 1,
\quad
c \equiv  \vert \tilde{\gamma} - \gamma + k_R^1 - k_R^2\vert.
\ee
\end{proposition}

\medskip

\noindent {\bf Remark:}
The non-negative integer coupling parameters $a, b, c$ that arise in this case satisfy
the condition $b \geq a + 1$.

\subsection{Case III: $m = n + 2$, $r = s = n + 1$, $N = 2 n + 2$}

Now the
fixpoint  subgroups of the
two different involutions
 $\sigma_L = \theta_{n + 1, n + 1}$ and
$\sigma_R = \theta_{n + 2, n}$ are
$U(N)_L \cong U(n + 1) \times U(n + 1)$
and
$U(N)_R \cong U(n + 2) \times U(n)$.
We consider the reductions associated with
 UIRREPS $(\rho,V)$ of $G$ (\ref{4.1}) having the form
\be
\rho := \left(
\rho^{(n + 1)}_{(k_L^1, 0)} \boxtimes \rho^{(n + 1)}_{(k_L^2, 0)}
\right)
\boxtimes
\left(
\rho^{(n + 2)}_{(k_R^1, a_1 \varpi_1)} \boxtimes \rho^{(n)}_{(k_R^2, 0)}
\right),
\label{4.84}\ee
where $a_1 \in \bZ_{\geq 0}$ and $k_L^1, k_L^2, k_R^1, k_R^2 \in \bZ$,
and the representation space is identified as $V \equiv V_{a_1 \varpi_1}^{(n + 2)}$.
Any
$X \in \cG$ is a pair $X = (X_L, X_R)$ with
$X_L \in \mfu(n + 1) \oplus \mfu(n + 1)$
and $X_R \in  \mfu(n + 2) \oplus \mfu(n)$,
and we may further write $X_L=(X_L^1, X_L^2)$ and  $X_R=(X_R^1, X_R^2)$,
where now $X_L^1, X_L^2 \in \mfu(n + 1)$, $X_R^1 \in \mfu(n + 2)$ and
$X_R^2 \in \mfu(n)$.
Then the $\cG$-representation  can be written as
\bea
&&\rho\prime(X) = \pi_{a_1 \varpi_1}^{(n + 2)} \left(
X_R^1 - \frac{\tr(X_R^1)}{n + 2} \1_{n + 2} \right) \nonumber\\
&&\phantom{X} + \left[k_L^1 \tr(X_L^1) + k_L^2 \tr(X_L^2)
+ \left( k_R^1 + \frac{\mu_{n + 2}(a_1 \varpi_1)}{n + 2} \right) \tr(X_R^1) + k_R^2 \tr(X_R^2)\right]\Id_V.
\label{rhoprime_III}
\eea
\begin{lemma}
The KKS ansatz (\ref{4.84}) yields admissible UIRREPS if and only if
$\exists\, \gamma, \tilde{\gamma}, {\hat{\gamma}} \in \bZ_{\geq 0}$
and $k \in \bZ$ such that the parameters $k_L^1, k_L^2, k_R^1, k_R^2 \in \bZ$
and $a_1 \in \bZ_{\geq 0}$ satisfy the conditions
\be
a_1 = \gamma n + \tilde{\gamma} + {\hat{\gamma}},
\quad
k_L^1 = k,
\quad
k_L^2 = \tilde{\gamma} - {\hat{\gamma}} + k,
\quad
k_R^1 = R - \tilde{\gamma} - k,
\quad
k_R^2 = {\hat{\gamma}} - \gamma - k,
\label{cond_III}
\ee
where $a_1 = Q + (n + 2) R$
with uniquely determined
$Q = Q(\gamma, \tilde{\gamma}, {\hat{\gamma}}) \in \{ 0, 1, \ldots, n + 1 \}$
and $R = R(\gamma, \tilde{\gamma}, {\hat{\gamma}}) \in \bZ$.
If the above conditions are met, then  $\dim(V^K) = 1$ and concretely
\be
V^K =
V_{a_1 \varpi_1}^{(n + 2)}[ \gamma e_1 + \gamma e_2 + \cdots + \gamma e_n
+ \tilde{\gamma} e_{n + 1} + {\hat{\gamma}} e_{n + 2}]
= \Span_\bC \{ \ket{ \gamma, \gamma, \ldots, \gamma, \tilde{\gamma}, {\hat{\gamma}} } \},
\ee
where the last equality refers to the bosonic oscillator realization of $V_{a_1 \varpi_1}^{(n + 2)}$.
\end{lemma}
\textbf{Proof.}
For the isotropy subalgebra we have
$\cK = \cM_\diag = \{ X = (X_0, X_0) \, | \, X_0 \in \cM \}$,
where
\be
\cM = \left\{
X_0 = \ri \begin{bmatrix}
D & 0 & 0 & 0\\
0 & \omega & 0 & 0 \\
0 & 0 & \tilde{\omega} & 0 \\
0 & 0 & 0 & D
\end{bmatrix}
\, \bigg| \,
D = \diag(d_1, d_2, \ldots, d_n) \in \bR^{n \times n},\, \omega, \tilde{\omega} \in \bR
\right\}.
\ee
Any $X=(X_L,X_R) \in \cK$ satisfies $X_L = X_R = X_0$, and therefore it has the components
\be
X_L^1 = \ri \begin{bmatrix}
D & 0 \\
0 & \omega
\end{bmatrix},
\quad
X_L^2 = \ri \begin{bmatrix}
\tilde{\omega} & 0 \\
0 & D
\end{bmatrix},
\quad
X_R^1 = \ri \begin{bmatrix}
D & 0 & 0 \\
0 & \omega & 0 \\
0 & 0 & \tilde{\omega}
\end{bmatrix},
\quad
X_R^2 = \ri D.
\ee
For any real $(n + 1)$-tuple $\varphi
= (\varphi_1, \varphi_2, \ldots, \varphi_{n + 1}) \in \bR^{n + 1}$
we let $\bar{\varphi} := \sum_{j = 1}^{n + 1} \varphi_j$,
$\tilde{\varphi} := \sum_{j = 1}^n \varphi_j$,
and introduce the traceless matrices
\be
H_\varphi := \diag(\varphi_1, \varphi_2, \ldots, \varphi_{n + 1}, - \bar{\varphi}),
\quad
H_R^2 := \diag(\varphi_1, \varphi_2, \ldots, \varphi_{n})
- \frac{\tilde{\varphi}}{n} \1_{n},
\ee
\be
H_L^1 := \diag(\varphi_1, \varphi_2, \ldots, \varphi_{n + 1})
- \frac{\bar{\varphi}}{n + 1} \1_{n + 1}, \quad
H_L^2 := \diag(- \bar{\varphi}, \varphi_1, \ldots, \varphi_{n})
+ \frac{\varphi_{n + 1}}{n + 1} \1_{n + 1}.
 \ee
We then write the components of $X\in \cK$ in the form
\be
 X_R^1 = \ri H_\varphi + \ri x \1_{n + 2},  \qquad
X_R^2 = \ri H_R^2 + \ri \left( x + \frac{\tilde{\varphi}}{n} \right) \1_n,
\ee
\be
 X_L^1 = \ri H_L^1 + \ri \left( x + \frac{\bar{\varphi}}{n + 1} \right) \1_{n + 1},
 \qquad
X_L^2 = \ri H_L^2 + \ri \left( x - \frac{\varphi_{n + 1}}{n + 1} \right) \1_{n + 1}.
\ee
From (\ref{rhoprime_III})
it follows that for any $ v \in V_{a_1 \varpi_1}^{(n + 2)}$  and $X\in \cK$ we have
\bea
&&\rho\prime(X) v
=  \pi_{a_1 \varpi_1}^{(n + 2)}(\ri H_\varphi) v
+ \ri (k_L^1 \bar{\varphi} - k_L^2 \varphi_{n + 1} + k_R^2 \tilde{\varphi}) v \nonumber \\
&&\phantom{XXX} + \ri x \left(
\mu_{n + 2}(a_1 \varpi_1) + (n + 2)k_R^1
+ (n + 1)(k_L^1 + k_L^2) + n k_R^2 \right) v.
\eea
Clearly $\rho\prime(X) v = 0$ $(\forall X \in \cK)$
if and only if
\be
\pi_{a_1 \varpi_1}^{(n + 2)}(H_\varphi) v
= (k_L^2 \varphi_{n + 1} - k_L^1 \bar{\varphi} - k_R^2 \tilde{\varphi}) v
\quad
(\forall \varphi \in \bR^n),
\label{Lcond1}\ee
and
\be
\mu_{n + 2}(a_1 \varpi_1) + (n + 2)k_R^1
+ (n + 1)(k_L^1 + k_L^2) + n k_R^2 = 0.
\label{Lcond2}\ee
Since
\be
k_L^2 \varphi_{n + 1} - k_L^1 \bar{\varphi} - k_R^2 \tilde{\varphi}
= - (k_L^1 + k_R^2) (e_1 + e_2 + \cdots + e_n)(H_\varphi)
+ (k_L^2 - k_L^1) e_{n + 1}(H_\varphi),
\ee
we obtain from (\ref{Lcond1}) that we must have
\be
V^K = V_{a_1 \varpi_1}^{(n + 2)}
[-(k_L^1 + k_R^2) (e_1 + e_2 + \cdots + e_n)
+ (k_L^2 - k_L^1) e_{n + 1}].
\label{4.98}\ee
It is easy to see  (cf.~Appendix A) that the weight space
in (\ref{4.98}) is non-trivial
if and only if $\exists\, (l_1, l_2, \ldots, l_{n + 2}) \in \bZ_{\geq 0}^{n + 2}$
with $l_1 + l_2 + \cdots + l_{n + 2} = a_1$, such that
\be
-(k_L^1 + k_R^2) (e_1 + e_2 + \cdots + e_n)
+ (k_L^2 - k_L^1) e_{n + 1}
= \sum_{j = 1}^{n + 1} (l_j - l_{n + 2}) e_j.
\label{4.99}\ee
We set
\be
\gamma := l_1,
\quad
\tilde{\gamma} := l_{n + 1},
\quad
{\hat{\gamma}} := l_{n + 2},
\quad
k:=k_L^1.
\ee
Then (\ref{4.99}) requires $l_1 = l_2 = \cdots = l_n = \gamma$ and
${\hat{\gamma}} - \gamma = k + k_R^2$
with $\tilde{\gamma} - {\hat{\gamma}} = k_L^2 - k$.
So, regarding $\gamma, \tilde{\gamma}, {\hat{\gamma}} \in \bZ$
and $k \in \bZ$ as \emph{free parameters},
we see that the other parameters have to obey the relations
\be
k_L^2 = \tilde{\gamma} - {\hat{\gamma}} + k,
\quad
k_R^2 = {\hat{\gamma}} - \gamma - k,
\quad
a_1 = \gamma n + \tilde{\gamma} + {\hat{\gamma}}.
\label{4.101}\ee
To satisfy the remaining condition (\ref{Lcond2}), we now
define $Q = Q(\gamma, \tilde{\gamma}, {\hat{\gamma}})
\in \{0, 1, \ldots, n + 1 \}$
and $ R = R(\gamma, \tilde{\gamma}, {\hat{\gamma}}) \in \bZ$
by the equality
\be
a_1 = \gamma n + \tilde{\gamma} + {\hat{\gamma}}
= Q + (n + 2) R.
\ee
Then (\ref{Lcond2}) translates into the  condition
$k_R^1 = R - \tilde{\gamma} - k$,
which completes the proof.
\hspace*{\stretch{1}} \qedsymb

\medskip
Further direct calculations  yield the explicit form of the reduced Laplacian (\ref{2.14}).
\begin{proposition}
Under the KKS ansatz (\ref{4.84}) parametrized by
arbitrary
$\gamma, \tilde{\gamma}, {\hat{\gamma}} \in \bZ_{\geq 0}$
and $k \in \bZ$ according to Lemma 4.6,
the reduced Laplacian of $U(N)$ satisfies
$- \Delta_{red} = H_{BC_n} + C$
with the constant
\be
C= -\frac{1}{6} n (4 n^2 + 12 n + 11)
+ \frac{1}{2} n (2 k + \tilde{\gamma} - {\hat{\gamma}})^2
+ (\tilde{\gamma} + k)(\tilde{\gamma} + k + 1)
+ ({\hat{\gamma}} - k)({\hat{\gamma}} - k + 1)
\ee
and coupling parameters given in the notation (\ref{**}) by
\be
a \equiv\gamma,
\quad
b \equiv\gamma + \tilde{\gamma} + 1,
\quad
c \equiv \gamma + {\hat{\gamma}} + 1.
\ee
\end{proposition}

\noindent {\bf Remark:}
The integer coupling parameters $a, b, c$  arising in this case satisfy
$ b, c \geq a + 1$.

\section{Discussion}
\setcounter{equation}{0}

We here summarize the results,  discuss the related work \cite{Obl} and
point out  open problems.

In this paper we applied the formalism of quantum Hamiltonian reduction under polar
group actions to study the reductions of the Laplace operator of $U(N)$
by means of the Hermann action (3.2) of the symmetry group
$G=(U(r) \times U(s)) \times (U(m) \times U(n))$
with $N=m + n=r+s$.
We concentrated on the 3 series of
cases for which the centralizer
of the corresponding section,
the group $K=M_{\diag}$ (\ref{3.41}), is Abelian.
We built the representation $(\rho,V)$ of the symmetry group that enters the definition of the reduction
by using as building blocks in  (\ref{4.2})  one-dimensional representations and a symmetric power of the
defining representation of the `largest' factor of $G$.
In the framework of this `KKS ansatz' we determined all cases for which the reduction is consistent
(that is $\dim(V^K) \neq 0$), and saw also that in these admissible cases
$\dim(V^K) =1$.
We then calculated the explicit formula of the reduced Laplacian by specializing equation (\ref{2.14}),
and found that up to an additive constant
it yields the $BC_n$ Sutherland Hamiltonian (\ref{**}) with coupling parameters given as follows:
\begin{itemize}
\item case I: $a, b, c \in \bZ_{\geq 0}$,
\item case II: $a, b, c \in \bZ_{\geq 0}$ with $b \geq a + 1$,
\item case III: $a, b, c \in \bZ_{\geq 0}$ with $b, c \geq a + 1$.
\end{itemize}
The dependence of the additive constant and of the coupling parameters $a, b, c$ on the
parameters of the respective representation $(\rho, V)$ is given by the 3 propositions formulated
in section 4.

The above results show that case I, which is the simplest case,
covers all  integral values of the coupling parameters $a,b,c$ and
the other two cases allow for alternative group theoretic descriptions of
the $BC_n$ model at proper subsets of the integral coupling parameters.
This state of affairs could not be foreseen before performing
the analysis of the different reduction schemes.
Observe also that if $b=c$, then the Hamiltonian (\ref{**}) becomes of type $C_n$, but
the $B_n$ and $D_n$ type Sutherland models do not arise from (\ref{**})
at any values of the integers $a$, $b$, $c$.
This is in  contrast with the corresponding classical Hamiltonian reduction
\cite{LMP}, which covers all coupling constants of the classical $BC_n$
model,  and is due to the never vanishing second term of the
 `measure factor' given by (\ref{mes}).
The measure factor represents  a kind of quantum anomaly  since
it gives the difference between the naive quantization of the reduced
classical Hamiltonian and the outcome of the corresponding
quantum Hamiltonian reduction  \cite{TMP}.

In case I, our analysis is consistent
with the results of Oblomkov \cite{Obl}, who studied
reductions
of the Laplace operator of $GL(m+n,\bC)$ using the symmetry group
\be
G^\bC:= (GL(m,\bC) \times GL(n,\bC)) \times (GL(m,\bC) \times GL(n,\bC)),
\qquad
m\geq n.
\label{5.1+}\ee
In fact, in case I our reduction is nothing but the compact real
form of the reduction studied in \cite{Obl} for $m=n$.
For the $m>n$ cases of the symmetry group (\ref{5.1+}) a generalization of the KKS ansatz
was  employed in \cite{Obl}, which  was found to yield
the complex version of the $BC_n$ Sutherland Hamiltonian (\ref{**}) with
integer coupling parameters subject to the restriction
 $c \geq b-(m-n) \geq 0$.  Thus
the coupling parameters obtained for $m>n$ form a proper subset of
those obtained for $m=n$,  and
this proper subset is different from those that we derived in our cases II and III.
For clarity we note that the KKS ansatz (\ref{KKSII})
that we adopted in case II was
motivated by the corresponding classical reduction  \cite{LMP}, and it
does not correspond to the ansatz used in \cite{Obl} for $m-n=1$.
It is not clear to us how the classical analogues
of the $m>n$ reductions of \cite{Obl} work.

Of course, the reductions
can be applied also to the differential operators associated with  the higher Casimirs.
This can be used
to explain the complete integrability of the $BC_n$ Sutherland model
and to derive the spectra as well as
the form of the joint eigenfunctions of the corresponding commuting Hamiltonians
at the pertinent values of the coupling constants from representation theory \cite{Obl}.

We stress  that the general method that we applied in our analysis
can be used also to study other problems in the future.
For example, one may try to determine all possible values
of the coupling constants  of the Sutherland models (\ref{1.1}) that may result
as reductions of the Laplacian of a compact Lie group in general.
This is closely related to the open problem concerning the classification
of the Hermann
actions and representations $(\rho, V)$ of symmetric subgroups $G$ (\ref{3.1}) such that the condition
$\dim(V^K)=1$
holds for the centralizer $K<G$ of the section.
In all such cases the reduced Laplace operator (\ref{2.14}) is expected to provide a
many-body model that can be solved by the group theoretic method
because of its very origin.

Besides the trigonometric real form that we considered,
the complex $BC_n$ Sutherland model admits the well known hyperbolic real form
and other physically very different real forms associated with
two types of particles \cite{Cal2,Hashi}.
The  derivation of the hyperbolic model by quantum Hamiltonian reduction can
be done similarly to the present work, but starting from $U(n,n)$
instead of $U(2n)$ (in case I) taking the Cartan involution both for $\sigma_L$ and for $\sigma_R$
(see also \cite{LMP}).
The models with two types of particles pose a more difficult problem.
At the classical level, it can be seen from \cite{Hashi} that to derive them
one needs to take the Cartan involution of $U(n,n)$ for $\sigma_L$
and a different involution for $\sigma_R$ that has a \emph{non-compact} fixpoint subgroup.
Therefore the corresponding quantum Hamiltonian reduction
would require some modifications of the
method used in this paper, which  need further investigation.

\newpage
\renewcommand{\thesection}{A}
\section{Some representation theoretic facts}
\renewcommand{\theequation}{A.\arabic{equation}}
\setcounter{equation}{0}

In this appendix we gather some basic facts in order
to fix the notations used in Section 4.

\subsection{On the UIRREPS of $SU(n)$ and $U(n)$}
Since the Lie group $SU(n)$ is compact,
connected and simply-connected, there is a
one-to-one correspondence between the UIRREPS
 $(\bfPi, V)$ of $SU(n)$  and the
finite dimensional complex IRREPS
$(\pi, V)$ of $\mfsl(n, \bC) = \mfsu(n)^\bC$.
In the complex simple  Lie algebra
$\mfsl(n, \bC)$
we have the Cartan subalgebra $\cH$ consisting of diagonal matrices, and use also
the real Cartan subalgebra
\be
\cH_\bR := \{ H \, | \, H \in \mfsl(n, \bC), \quad
H \mbox{ is diagonal with real entries} \} \subset \cH.
\ee
The functionals
$\{ e_i \}_{i = 1}^n \subset \cH^*$ are defined by the formula
$e_i(H) := H_{i i}$  $(H \in \cH)$.
The roots with respect to $\cH$ form the set
$\cR := \{ e_i - e_j \, | \, 1 \leq i, j \leq n,\,  i \neq j \} \subset \cH^*$
and we fix the root vectors $E_{e_i - e_j} := E_{i j}$.
The set of positive roots  is
$\cR_+ := \{ e_i - e_j \, | \, 1 \leq i < j \leq n \}$
and the  simple roots are $\alpha_i := e_i - e_{i + 1}$ $(1 \leq i \leq n - 1)$.
Let $\varpi_i = \sum_{k = 1}^i e_k \in \cH^*$ $(1 \leq i \leq n - 1)$
denote the fundamental weights.
The equivalence
classes of the IRREPS
of $\mfsl(n, \bC)$ can be uniquely
labeled by the highest
(dominant integral) weights, which  are the elements of
\be
P_+(SU(n)) = \{ a_1 \varpi_1 + a_2 \varpi_2 + \cdots + a_{n - 1} \varpi_{n - 1}
\, | \,
a_1, a_2, \ldots, a_{n - 1} \in \bZ_{\geq 0} \}
\cong \bZ_{\geq 0}^{n - 1}.
\label{P+}\ee
Now take an $\mfsl(n, \bC)$ IRREP $(\pi_\lambda, V_\lambda)$ of
highest weight $\lambda \in P_+(SU(n))$.
To any linear functional $\nu \in \cH^*$ we associate
the weight space
\be
V_\lambda [\nu] := \bigcap_{H \in \cH}
\ker \left(  \pi_\lambda(H) - \nu(H) \Id_{V_\lambda} \right) \subset V_\lambda,
\ee
and we also define the set of weights
$\cW_\lambda := \{ \nu \, | \, \nu \in \cH^*, V_\lambda[\nu] \neq \{ 0 \} \}$.
Then we have the weight space decomposition
$V_\lambda = \bigoplus_{\nu \in \cW_\lambda} V_\lambda[\nu]$.
Note that $\lambda \in \cW_\lambda$ and $\dim(V_\lambda[\lambda]) = 1$, so
we can write $V_\lambda[\lambda] = \bC v_\lambda$  with some
 highest weight vector $v_\lambda$.
The characteristic property of the non-zero vector $v_\lambda$ is that
$\pi_\lambda(E_{\alpha}) v_\lambda = 0$ holds for all $\alpha \in \cR_+$.
The IRREP $(\pi_\lambda, V_\lambda)$ of $\mfsl(n, \bC)$ induces the
UIRREP $(\bfPi_\lambda, V_\lambda)$ of $SU(n)$ by the
requirement
$\bfPi_\lambda(e^X) = e^{\pi_\lambda(X)}$ for all $X \in \mfsu(n)$.
The corresponding scalar product on $V_\lambda$  can be defined by fixing the norm of $v_\lambda$ and
requiring the anti-hermiticity of $\pi_\lambda(X)$ for all $X\in \mfsu(n)$.

The UIRREPS
of $U(n)$ are usually parametrized by the set
\be
P_+(U(n))= \{ \mathbf{m} =
(m_1,m_2, \ldots, m_n) \in \bZ^n\,\vert\, m_1 \geq m_2 \geq \cdots \geq m_n\}.
\ee
The representation $\rho_{\mathbf{m}}$ of $U(n)$  may be defined as the extension
of the representation $\bfPi_\lambda$ of $SU(n)< U(n)$
characterized by the properties
\be
\lambda = \sum_{i=1}^{n-1} (m_i- m_{i+1}) \varpi_i
\qquad\hbox{and}
\qquad
\rho_{\mathbf{m}}(\xi \1_n) = \xi^{m_1+ \cdots + m_n} \Id_{V_\lambda}
\qquad \forall \xi\in U(1).
\label{A.35}\ee
In the main text we use a slightly different parametrization by pairs
$(k, \lambda)\in \bZ \times P_+(SU(n))$.
The correspondence is given by the relation $m_1+\cdots + m_n = \mu_n(\lambda) + k n$,
as is seen from the comparison between  (\ref{A.35}) and  (\ref{mu_n_value}) and (\ref{Urep}).

\subsection{On the bosonic oscillator realization of
$(\pi_{m \varpi_1}, V_{m \varpi_1})$}
Fix an integer $n \geq 2$ and to each $n$-tuple
$(l_1, l_2, \ldots, l_n) \in \bZ_{\geq 0}^n$  associate a
`symbol' $\ket{l_1, l_2, \ldots, l_n}$. Let $\cF$ denote the
complex vector space generated by these symbols,
\be
\cF :=
\bigoplus_{(l_1, l_2, \ldots, l_n) \in \bZ_{\geq 0}^n}
\bC \ket{l_1, l_2, \ldots, l_n}.
\ee
Endow $\cF$ with the scalar product $(\, , \,)$ for which
the vectors
$\{\ket{l_1, l_2, \ldots, l_n}\}_{(l_1, l_2, \ldots, l_n) \in \bZ_{\geq 0}^n} $
satisfy
\be
\left( \ket{l_1, l_2, \ldots, l_n}, \ket{l'_1, l'_2, \ldots, l'_n} \right)
= \delta_{l_1, l'_1} \delta_{l_2, l'_2} \cdots \delta_{l_n, l'_n},
\ee
and introduce the  annihilation and creation operators
$b_i$ and $b_i^\dagger$ $(1 \leq i \leq n)$ on $\cF$ by
\begin{align}
& b_i \ket{l_1, l_2, \ldots, l_n}
:= \begin{cases}
\sqrt{l_i} \ket{l_1, l_2, \ldots, l_i - 1, \ldots, l_n} & \text{if $l_i \geq 1$}, \\
0 & \text{if $l_i = 0$},
\end{cases}
\\
& b_i^\dagger \ket{l_1, l_2, \ldots, l_n}
:= \sqrt{l_i + 1} \ket{l_1, l_2, \ldots, l_i + 1, \ldots, l_n}.
\end{align}
Then $b_i^\dagger$ is the adjoint of $b_i$, and one has the
commutation relations
\be
[ b_i, b_j] = 0,
\quad
[ b_i^\dagger, b_j^\dagger ] = 0,
\quad
[ b_i , b_j^\dagger ] = \delta_{i, j} \Id_{\cF}.
\ee
The `bosonic Fock space'  $\cF$
decomposes as the orthogonal direct sum
$\cF = \bigoplus_{m \in \bZ_{\geq 0}} \cF_m$
with
\be
\cF_m := \Span_\bC \{ \ket{l_1, l_2, \ldots, l_n}
\, | \,
(l_1, l_2, \ldots, l_n) \in \bZ_{\geq 0}^n,
\quad
l_1 + l_2 + \cdots + l_n = m \}.
\ee

Now consider the linear map $\psi \colon \mfgl(n, \bC) \rightarrow \End(\cF)$
defined on the standard basis
$\{ E_{ij} \}_{1 \leq i, j \leq n}$
of $\mfgl(n, \bC)$ by
\be
\psi(E_{ij}) := b_i^\dagger b_j.
\ee
Then $(\psi, \cF)$ is a representation of
$\mfgl(n, \bC)$ and the subspace $\cF_m$ is invariant under  $\psi$.
The map
\be
\psi_m \colon \mfgl(n, \bC) \rightarrow \End(\cF_m),
\quad
X \mapsto \psi_m (X) := \psi(X) |_{\cF_m}
\ee
provides a  finite dimensional
representation of the Lie algebra $\mfgl(n, \bC)$.
By restricting $\psi_m$ to the subalgebra
$\mfsl(n, \bC) < \mfgl(n, \bC)$, we end up with a
finite dimensional representation $(\psi_m, \cF_m)$ of
$\mfsl(n, \bC)$.
The set of weights of the representation $(\psi_m, \cF_m)$ is
\be
\cW_m := \left\{ \sum_{i = 1}^n l_i e_i
\, \bigg| \,
(l_1, l_2, \ldots, l_n) \in \bZ_{\geq 0}^n,
\quad
l_1 + l_2 + \cdots + l_n = m \right\},
\ee
and the weight space $\cF_m[\nu] \subset \cF_m$ corresponding
to weight $\nu = \sum_{i = 1}^n l_i e_i \in \cW_m$ takes the form
\be
\cF_m\left[ l_1 e_1 + l_2 e_2 +\cdots + l_n e_n \right] = \bC \ket{l_1, l_2, \ldots, l_n}.
\ee
Note that each weight space is one-dimensional.
The representation  $(\psi_m,\cF_m)$ contains the (up to rescaling)  unique highest weight
vector
$v_m := \ket{m, 0, \ldots, 0}$,
with weight $m\varpi_1 =m e_1 \in \cW_m$.
This shows that $(\psi_m, \cF_m)$
is equivalent to the IRREP $(\pi_{m \varpi_1}, V_{m \varpi_1})$.
We identify these  $\mfsl(n, \bC)$ (and the naturally corresponding $\mfsu(n)$)
 representations in the proofs presented in Section 4.

\medskip
\bigskip
\noindent{\bf Acknowledgements.}
We thank J.~Balog for useful comments on the manuscript.
This work was supported
by the Hungarian
Scientific Research Fund (OTKA) under the grant K 77400.

\end{document}